\documentclass[a4paper,11pt]{memoir}

\usepackage[english]{babel}
\usepackage[utf8x]{inputenc}
\usepackage[T1]{fontenc}

\usepackage[a4paper,top=3cm,bottom=2cm,left=3cm,right=3cm,marginparwidth=1.75cm]{geometry}

\usepackage{blindtext}
\usepackage{algorithm}
\usepackage{algorithmic}
\usepackage{amsmath}
\usepackage{amsthm}
\usepackage{amssymb}
\usepackage{graphicx}
\usepackage{slashbox}
\usepackage[colorinlistoftodos]{todonotes}
\usepackage[colorlinks=true, allcolors=blue]{hyperref}
\usepackage{mathtools}
\usepackage{multirow}
\usepackage{MnSymbol}
\usepackage{relsize}

\newtheorem{deff}{Definition}[section]
\newtheorem{preuve}{Proof}[section]
\newtheorem{theorem}{Theorem}[section]
\newtheorem{prop}{Proposition}[section]
\newtheorem{exemple}{Example}[section]
\newtheorem{remark}{Remark}[section]
\newtheorem{claim}{Claim}[section]
\newtheorem{corollary}{Corollary}[section]
\newtheorem{lemma}{Lemma}[section]

\newcommand*{\LargerCdot}{\raisebox{-0.25ex}{\scalebox{1.2}{$\cdot$}}}
\newcommand\numberthis{\addtocounter{equation}{1}\tag{\theequation}}
\usepackage{pdfpages}

\begin{document}
\includepdf[pages={1}]{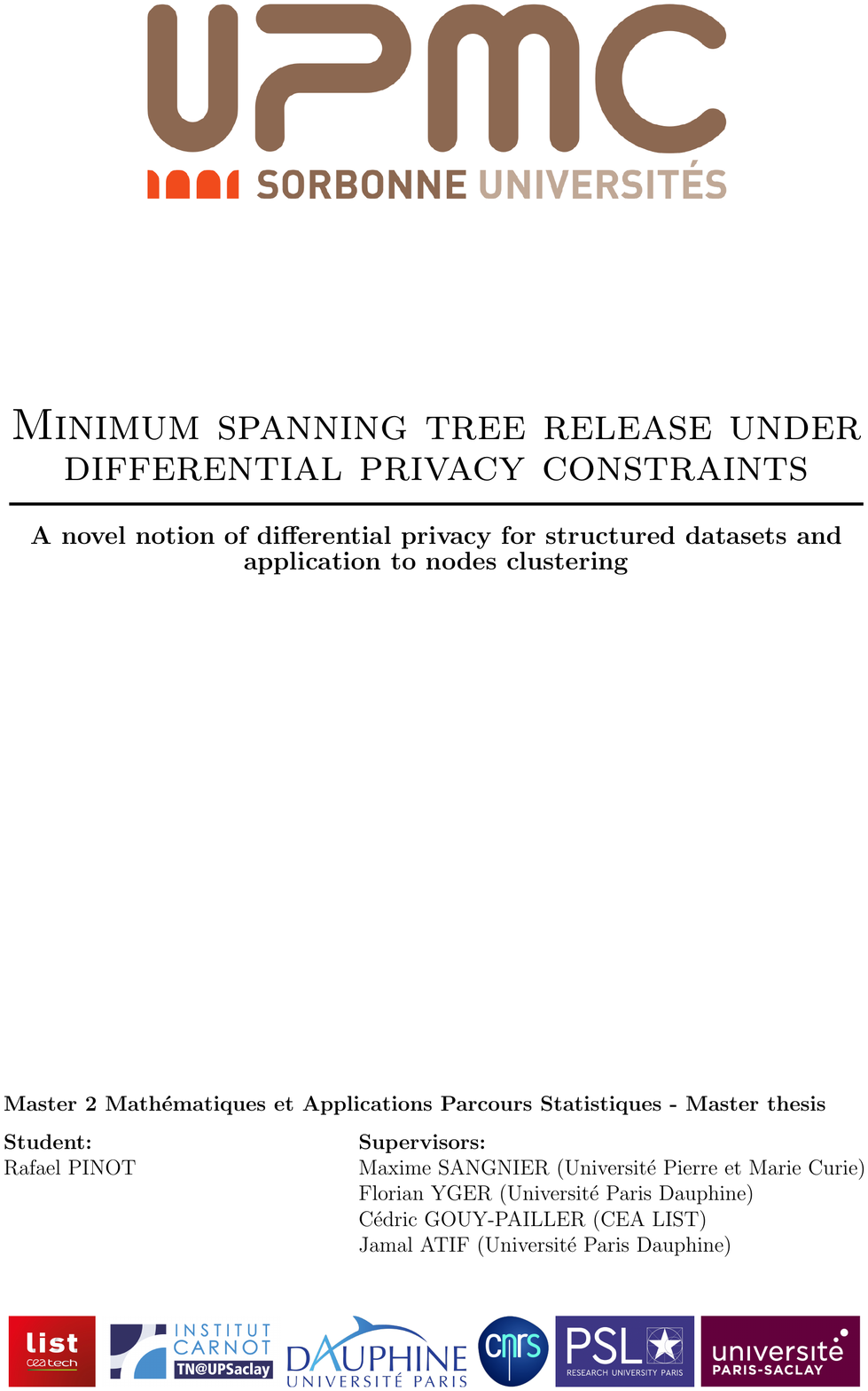}
\begin{minipage}{0.85\textwidth}

\section*{\hspace*{\fill} Abstract \hspace*{\fill}}
We investigate the problem of nodes clustering under privacy constraints when representing a dataset as a graph. Our contribution is threefold. First we formally define the concept of differential privacy for structured databases such as graphs, and give an alternative definition based on a new neighborhood notion between graphs. This definition is adapted to particular frameworks that can be met in various application fields such as genomics, world wide web, population survey, etc. 
Second, we introduce a new algorithm to tackle the issue of privately releasing an approximated minimum spanning tree topology for a simple-undirected-weighted graph. It provides a simple way of producing the topology of a private almost minimum spanning tree which outperforms, in most cases, the state of the art "Laplace mechanism" in terms of weight-approximation error.
A thorough theoretical analysis of our algorithm stressing the comparison between our bound and the state of the art theoretical bound is presented. Given a simple-undirected-weighted graph $G=(V,E,w)$, our algorithm will always outperform the "Laplace mechanism" when $|E|\geq 2(|V|-1)$. To illustrate and support this result we also perform some experiments comparing the two methods based on simulated graphs.

Finally, we propose a theoretically motivated method combining a sanitizing mechanism (such as Laplace or our new algorithm) with a Minimum Spanning Tree (MST)-based clustering algorithm. It provides an accurate method for nodes clustering in a graph while keeping the sensitive information contained in the edges weights of the private graph. We provide some theoretical results on the robustness of an almost minimum spanning tree construction for Laplace sanitizing mechanisms. These results exhibit which conditions the graph weights should respect in order to consider that the nodes form well separated clusters both for Laplace and our algorithm as sanitizing mechanism. The method has been experimentally evaluated on simulated data, and preliminary results show the good behavior of the algorithm while identifying well separated clusters.

\vspace{1cm}
\section*{\hspace*{\fill} Acknowledgements \hspace*{\fill}}

I would like to express my gratitude to the \textsc{LADIS} and the \textsc{LAMSADE} laboratories for hosting me during this internship. I would also like to thank my supervisors for their guidance and encouragement throughout this work. Last but not least, I give thanks to all the researchers I had the opportunity to meet, especially Anne MORVAN and Arnaud GRIVET-SÉBERT for all the helpful discussions we had.
\end{minipage}
\newpage
  
\vspace*{\fill}
\tableofcontents
\vspace*{\fill}

\vspace*{\fill}
\chapter*{Introduction}
\addcontentsline{toc}{chapter}{Introduction}
Graphs represent a useful representation for many types of data, widely used in e.g. bioinformatics, network analysis, etc. 
More broadly, any dataset can be converted into a graph by using a well-chosen similarity matrix construction. 
In that respect, graph clustering~\cite{Schaeffer07} appears to be a key tool for understanding the underlying structure of many data sets by locating nodes groups ruled by a specific similarity.

A primary issue to be tackled while using machine learning techniques on a dataset is to protect the private characteristics of the individuals belonging to this dataset. Protecting the privacy of individuals while producing statistical analysis is an old topic which foundations have mostly been laid in the 1980's by Goldwasser and al., Denning or Adam and al. in~\cite{GOLDWASSER1984270,Denning1980,Adam1989}. A very famous example of massive security breach due to data-analysis, "the Netflix Prize contest" ~\cite{Narayanan_2008} brought this concern back in light by showing that it is possible to obtain a robust de-anonymization of a database by cross correlating the anonymized database with some other open source datasets. Following this, recent studies including ~\cite{Fredrikson_2014} and ~\cite{Fredrikson_2015} revealed that our classical methods for data analysis are vulnerable to reconstruction attacks. It means that, given some statistics about a dataset, an adversarial machine learning algorithm can obtain with high probability personal characteristics from the individuals in the dataset. This might raise serious issues in medical applications for instance~\cite{Fredrikson_2014}.
Those concerns are real, and several definitions and applications of what privacy should mean and be used for have been introduced. For a complete survey on the definitions of Privacy-preserving data publishing one could refer to~\cite{Fung:2010:PDP:1749603.1749605}. The widely adopted definition of what a "good" privacy condition should be is  called "differential privacy".
Since this seminal work~\cite{Dwork_2006}, several models respecting the differential privacy conditions have been proposed (e.g.~\cite{Zhanglong_Ji_2014}), but most of them do not consider structured types of data such as graphs. 
Therefore, how to consider differential privacy on structured data types remains an open question. Should one keep private the nodes, the edges and/or the weights of the graph. Those choices are mainly made according to the nature of the data at hand.

This work aims at investigating formal frameworks for preserving privacy in  learning from graph-structured data. The overall goal is to pave the way for clustering applications in genomics, proteomics, web monitoring, etc. where privacy-preserving is not an option but a strong requirement. In the sequel, after introducing what one should understand by differential privacy on structured databases, we will focus on providing new ways to analyze those datasets while preserving privacy in the process. Chapters~\ref{section:IntroToDP} and~\ref{section:DPforGraphs} provide introductions to differential privacy and differential privacy on structured datasets. Chapter~\ref{section:PrivateMSTs} presents our contribution to the theory of differential privacy as well as a new algorithm for the release of a minimum spanning tree under privacy conditions and compare it to the state of the art for releasing such a statistic. Finally, Chapter~\ref{section:clustering} introduces our contribution to differentially private clustering while presenting our first advances on MST-based differentially private clustering. Since this work does not recall the foundations of graph theory, we recommend reading appendix~\ref{appendix:graphnotation} to ensure the readability of the following report.
\vspace*{\fill}

\chapter{An introduction to differential privacy}
\label{section:IntroToDP}

Chapter~\ref{section:IntroToDP} aims at providing a simple but efficient introduction to the notions and preliminaries of differential privacy. They will first be presented in an intuitive way, and further formally stated. Some classical results will also be presented. To conclude this chapter we introduce the Exponential mechanism which is the cornerstone of the differentially private analysis for structured databases.

\section { Differential privacy Framework}
\begin{figure}[!ht]
\centering
\includegraphics[width=0.8\textwidth]{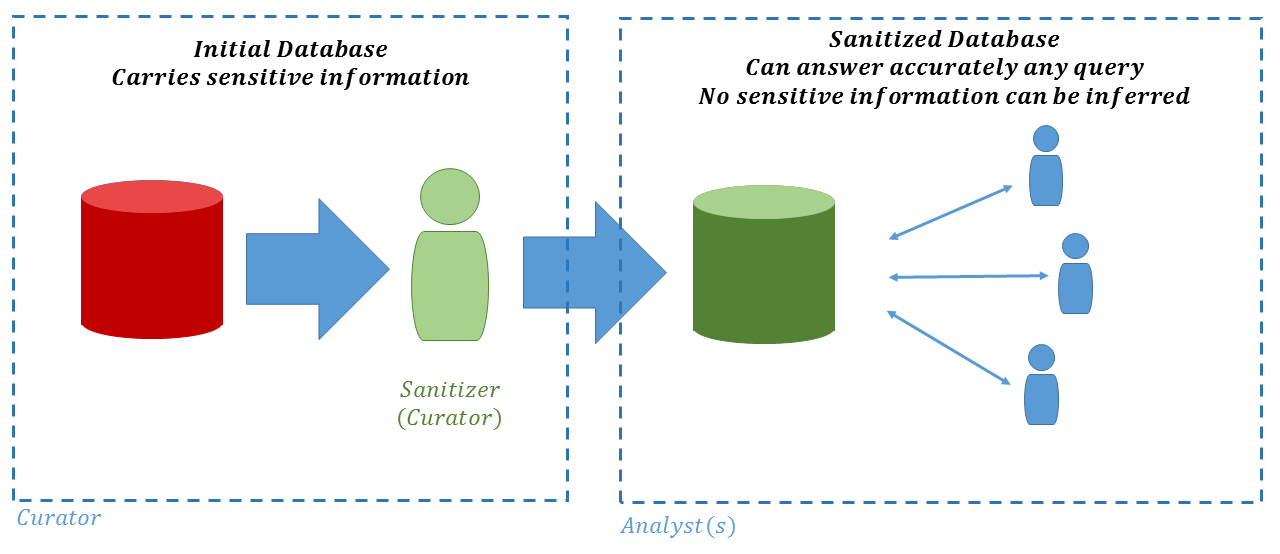}
\caption{The Sanitizing Pipe Dream}
\label{figure:DreamPipe}
\end{figure}

The Framework for differential privacy inherits from a long history of formalizing privacy concerns which here, for simplicity, is summed up to two settings presentation including the ideal framework and the actually used one.
The first assumption in the differential privacy framework is that an agent, called curator holds the database. This curator is trustworthy (will not leak the database), and has sufficient (unlimited) computational power to protect the database from being hacked. Moreover, the database is in a safe place and can not be physically stolen. Those assumptions essentially allows one to consider that the only way to have access to the database is through the curator. Therefore, an adversarial analyst can only infer information about the individual in the database by statistical reconstruction, based on the information released by the curator. Finally to cover a wide range of settings, one can consider the adversarial analyst to have unlimited computational power and unlimited knowledge (i.e apart from what he/she is trying to infer, he/she has access to any kind of additional databases).

\begin{figure}[!ht]
\centering
\includegraphics[width=0.85\textwidth]{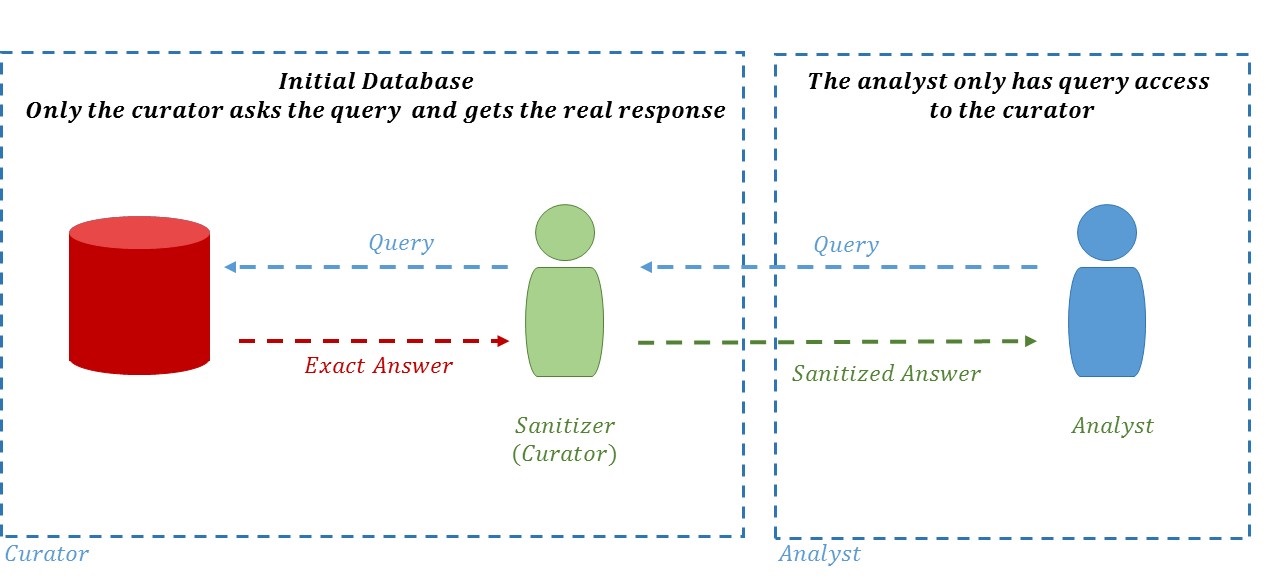}
\caption{Actual framework for differentially private query answering}
\label{figure:Realpipe}
\end{figure}

Figure~\ref{figure:DreamPipe} represents the ideal version of privacy preserving data analysis as first presented by Cynthia Dwork~\cite{Dwork_2013}. In this so called "Sanitizing Pipe dream", the curator can released a new database that could hypothetically accurately answer any kind of query without leaking any private information about the database. Of course, and as Claim~\ref{Claim:TheFundamentalLawInformationRecovery} will formalize it later, this version is not applicable while preserving differential privacy.
In practice, as illustrated by Figure~\ref{figure:Realpipe}, the analyst only has query access to the curator and get sanitized response instead of real query answers. In the remaining of this report, every mechanism presented to ensure differential privacy is applied by the curator to answer the corresponding query asked by the analyst.
Section~\ref{section:understandprivacy} aims at giving a good intuition of why the curator is crucial to keep the sensitive information in the database private.

\section { How should one understand differential privacy}
\label{section:understandprivacy}
First introduced by Dwork and al. in~\cite{Dwork_2006}, differential privacy is a robust and rigorously defined notion that provides formal but adaptive conception of privacy preserving data-analysis, but let us first give an intuitive understanding of it. The main idea in conceiving such a privacy is that any individual information is protected if \textit{"the outcome of any analysis is essentially equally likely, independent of whether any individuals joins, or refrains from joining the database"}.
Let us clarify this statement with some examples.

\begin{exemple}[Differencing attack]
\label{Example:MrPresident}
Assume some adversarial analyst has a query access to a medical record database containing the French President's record without any curator being set. He/she wants to know if the president has some specific medical condition, let it be a heart disease, but he/she is only allowed to produce some statistical analysis, e.g the frequency of this condition in the database (only has query access to the database). Since, given a database, the analysis outcome is deterministic (frequency),  any outcome will not be essentially equally likely to occur if one individual refrains from joining the database. The differencing attack (cf. Figure~\ref{figure:DefferencingAttack}) uses this flaw to infer some extremely sensitive information about Mr President. 

\begin{figure}[!ht]
\centering
\includegraphics[width=\textwidth]{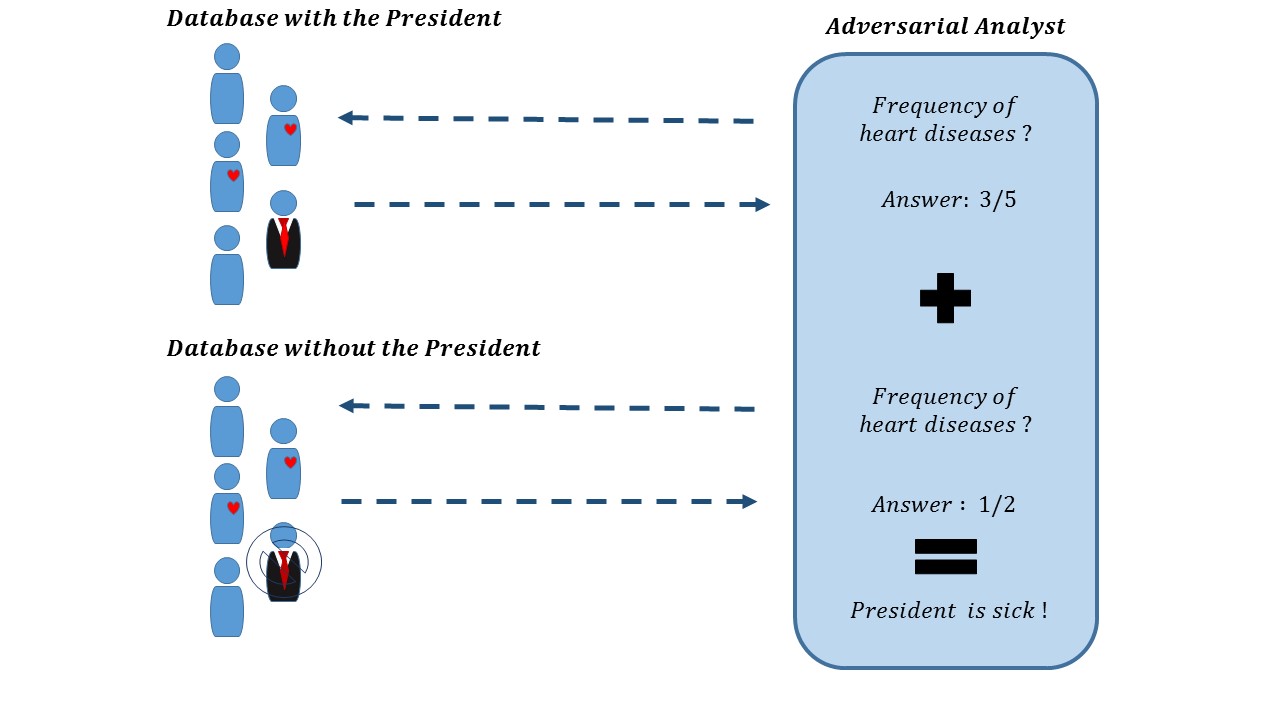}
\caption{Inferring the French President medical condition while asking the same query on two adversarial chosen databases}
\label{figure:DefferencingAttack}

\end{figure}

As illustrated by figure~\ref{figure:DefferencingAttack} differencing attack consists in producing the analysis, first on the initial database containing the president, and second on the same database without the president being in the studied population. Assuming the analyst knows the size of the database, the output variation between the two analysis gives with perfect certainty the medical condition of the French President.
\end{exemple}

Example~\ref{Example:MrPresident} highlights the fact that any deterministic analysis of a dataset represents a security breach, this is why the curator is needed to input some kind of randomness that will help preserving privacy. However, it does not mean that any statistical analysis is prohibited, it only means that perfect accuracy breaks privacy.
Example~\ref{Example:JohnDoe} represents a toy setting where an analyst can produce a useful statistical analysis on a given population with no privacy breach on any individual.
\begin{exemple}
\label{Example:JohnDoe}
Let us suppose any analyst wants to know how much eyes does the average American have. Given a representative dataset for the American population, any simple analysis will give with high probability that the average American has two eyes. Let us now suppose that an individual, named John Doe, was part of the dataset used for this analysis. \\ Was the analysis a security breach of John Doe's information according to differential privacy?
No, it was not! In fact, whether John participates or not to that kind of study will not significantly change it outcomes (the average American will have two eyes either way), and that is what differential privacy is about. Therefore the analyst just produced a useful analysis allowing him/her to learn about the American population without breaking the private information of John Doe.
\end{exemple}

\begin{remark}
This definition of privacy is not too permissive since considering Example~\ref{Example:JohnDoe} as a privacy breach rules out any statistical analysis of a dataset. 
\end{remark}

Section~\ref{section:FormalDefinition} provides a formal definition of differential privacy. This definition is closely related to the notion of randomness to ensure statistical indistinguishableness of the analysis outcomes. Statistical indistinguishableness also refers to one of the most important notion in differential privacy, the neighborhood notion. 

\section{Randomness and Neighboring}
\label{section:FormalDefinition}

To provide a formal definition of differential privacy, one should first better introduce the notions of randomness, and neighborhood. Randomness is essential because deterministic algorithms represents privacy breaches as illustrated by Example~\ref{Example:MrPresident}. It should be more convenient to preserve some kind of accuracy and ensure the statistical indistinguishableness of the outcomes. Moreover, giving a good definition of two statistically similar outcomes is not trivial. Intuitively, two outcomes are statistically close if they are computed on the same kind of databases, and if they are close in the output space. This is why one 
also needs to well introduce the notion of neighboring between databases and between output distributions.

\begin{deff}[Randomized Algorithm]
Given a probability space $(\Omega,\mathcal{F},\mathbb{P})$, and a measurable set $B$, a randomized algorithm $\mathcal{A}$ with domain $D$ and range $B$ is a function that takes an input $ d \in D$, and outputs a random variable (denoted $\mathcal{A}(d)$) with values in $B$.
\end{deff}

We will always consider for the remaining of this work the underlying probability space $(\Omega,\mathcal{F},\mathbb{P})$, denote $Range(\mathcal{A})$ the output space $B$, and omit to identify it when the setting is clear enough. 
Moreover we simplify the database notation. One should think of a database $x$ as being collections of records from a universe $\mathcal{X}$. For convenience, databases will often be represented by their histograms: $x \in \mathbb{N}^{|\mathcal{X}|}$, in which (by notation-abuse) for all $ i \in \mathcal{X}$, $x_{i}$ represents the number of elements of type $i$ in $x$. In this representation, a natural measure  of the distance between two databases $x$ and $y$ will be their $\ell_{1}$ distance:

\begin{deff}[Distance between two databases] The $\ell_{1}$ norm of a database $x$ is denoted $||x||_{1}$ and defined to be:
$$||x||_{1} = \sum\limits_{i=1}^{|\mathcal{X}|}|x_{i}|$$
The $l_{1}$ distance between two databases $x$ and $y$ is $||x-y||_{1}$. Moreover we call neighboring databases any $x,y \in \mathbb{N}^{|\mathcal{X}|}$ such that $||x-y||_{1} \leq 1$ (denoted $x \sim y$).
\end{deff}

The neighborhood notion between databases is set, let us now give such a notion for the output distributions. It is based on the well known Kullback-Leibler divergence.

\begin{deff}[Kullback-Leibler Divergence]
The Kullback-Leibler Divergence, or relative entropy, between two random variables $Y$ and $Z$ taking values from the same domain is defined to be :
$$D\left(Y||Z\right)=\mathbb{E}_{y\sim Y}\left[\ln\frac{\mathbb{P}(Y=y)}{\mathbb{P}(Z=y)} \right].$$
\end{deff}

The Kullback-Leibler divergence is not a distance since it does not respect the triangle inequality and is not symmetric, however, it provides a good intuition of one-sided neighborhood for statistical distributions.
Definition~\ref{definition:MaxDivergence} is a generalization of the Kullback-Leibler Divergence which is closely related to the definition of differential privacy as stated in proposition~\ref{proposition:KullbackisDifferentialprivacy}.

\begin{deff}[Max Divergence]
\label{definition:MaxDivergence}
The Max divergence between two random variables $Y$ and $Z$ taking values from the same domain is defined to be: 
$$ D_{\infty}\left(Y||Z\right)=\sup\limits_{S \subset Supp(Y)}\left[\ln\frac{\mathbb{P}(Y\in S)}{\mathbb{P}(Z\in S)} \right]. $$

The $\delta$-Approximate Max divergence between Y and Z is defined to be

$$ D^{\delta}_{\infty}\left(Y||Z\right)=\sup\limits_{S \subset Supp(Y) \textit{ and } \mathbb{P}(Y\in S) > \delta }\left[\ln\frac{\mathbb{P}(Y\in S) - \delta}{\mathbb{P}(Z\in S)} \right]. $$
\end{deff}

Let us now introduce the formal definition of differential privacy, and demonstrate how $\delta$-Approximate Max divergence and differential privacy notions overlap.

\begin{deff}[Differential Privacy]
A randomized algorithm $\mathcal{A}$ with domain $\mathbb{N}^{|\mathcal{X}|}$ is called $(\epsilon,\delta)$-differentially private if for any $S \subset Range(\mathcal{A})$ and for all $x,y \in \mathbb{N}^{|\mathcal{X}|}$ such that $||x-y||_{1} \leq 1 $: $$\mathbb{P}\left[\mathcal{A}(x) \in S\right] \leq e^{\epsilon}\mathbb{P}\left[\mathcal{A}(y) \in S\right] + \delta $$

If $\delta=0$, $\mathcal{A}$ is said to be $\epsilon$-differentially private.
\end{deff}

This formal definition provides a guarantee that under a reasonable choice of the privacy degree $\epsilon$, two close databases in term of $\ell_{1}$ norm will produce statistically inseparable results. It is also a symmetrical rewriting of $\delta$-Approximate Max divergence.

\begin{prop}[Privacy by Kullback-Leibler Divergence]
\label{proposition:KullbackisDifferentialprivacy} A randomized algorithm $\mathcal{A}$ is $(\epsilon,\delta)$-differentially private if and only if for all $x,y \in \mathbb{N}^{|\mathcal{X}|}$ such that $x \sim y$,  $D^{\delta}_{\infty}\big(\mathcal{A}(x)||\mathcal{A}(y)\big) \leq \epsilon$ and $D^{\delta}_{\infty}\big(\mathcal{A}(y)||\mathcal{A}(x)\big) \leq \epsilon$
\end{prop}

\begin{preuve}[Sketch of proof]
Let $\mathcal{A}$ be $(\epsilon,\delta)$-differentially private randomized algorithm. Let $S \subset Range(\mathcal{A})$ and $x \sim y \in \mathbb{N}^{|\mathcal{X}|}$, the proof comes down to this equivalence \begin{align*}
\mathbb{P}[\mathcal{A}(x) \in S] \leq e^{\epsilon}\mathbb{P}[\mathcal{A}(y) \in S] + \delta &\iff \mathbb{P}[\mathcal{A}(x) \in S] - \delta \leq e^{\epsilon}\mathbb{P}[\mathcal{A}(y) \in S] \\
&\iff \ln \frac{\mathbb{P}[\mathcal{A}(x) \in S] - \delta}{\mathbb{P}[\mathcal{A}(y) \in S]} \leq \epsilon 
\end{align*}
for all $S$ such that $\mathbb{P}[\mathcal{A}(x) \in S] > \delta$, $\mathbb{P}[\mathcal{A}(y) \in S] > 0$ . \hfill\ensuremath{\blacksquare}

\end{preuve}

\begin{remark}
For a full proof one can refer to appendix~\ref{appendix:proofKulback}.
\end{remark}

The definition of differential privacy is actually a characterization of a privacy preserving algorithm, it is thus a property some mechanisms should help preserving. Section~\ref{section:ClassicalResults} provides some of the most famous and used mechanisms that aim at sanitizing a statistical analysis without degrading too much its accuracy.

\section{Classical results}
\label{section:ClassicalResults}
Section~\ref{section:ClassicalResults} introduces the most classical results about differential privacy, namely: the Laplace mechanism, the Gaussian mechanism and the composition theorems. 

Let us consider that a numeric query is a deterministic function $f:\mathbb{N}^{|\mathcal{X}|} \rightarrow \mathbb{R}^k$. This kind of queries maps a database to a $k$-dimensional real vector. In the sequel we will denote numerical query as query, and specify only when it is not numerical. One of the most important notion in differential privacy to measure the accuracy of such queries is their sensitivity. The very first work by Dwork, McSherry, Nissim, and Smith~\cite{Dwork_2006} on differential privacy introduces the idea that the sensitivity of a query is the measure of how much, in the worst case, a single individual's data can change the answer to the query. In their~\cite{Nissim_2007}, Nissim and al. give an extended theoretical study on queries sensitivity and its link to elaborating differentially private mechanisms. Since then, most of differential privacy studies rely on this notion. 

\begin{deff}[$\boldsymbol{\ell_{p}}$ sensitivity]
For any $p \in \mathbb{N}$, and $f: \mathbb{N}^{|\mathcal{X}|} \rightarrow \mathbb{R}^{k}$, the $\ell_{p}$ sensitivity of f is:  $$ \Delta_{p}(f) := \sup\limits_{x,y \in \mathbb{N}^{|\mathcal{X}|}, x \sim y} || f(x) - f(y) ||_{p}$$ i.e the maximum $\ell_{p}$-distance between the outcomes of any two neighboring databases.
\end{deff}

This definition is mostly used in its $\ell_{1}$ and $\ell_{2}$ form, one can respectively use those definitions to present the Laplace and Gaussian mechanism. 

\begin{deff}[Laplace Mechanism]
Given any function $f: \mathbb{N}^{|\mathcal{X}|} \rightarrow \mathbb{R}^{k}$, and $ \epsilon > 0$, the Laplace mechanism is defined as 
$$ \mathcal{M}_{L}(x,f,\epsilon)= f(x) + \left(Y_{1},...,Y_{k}\right)$$
where $Y_{i}$ are i.i.d. random variables drawn from $Lap\left( \Delta_{1}(f)/\epsilon \right)$.

Where $Lap(b)$ denote the Laplace distribution with scale $b$ that is the distribution with probability density  $Lap(x|b)=\frac{1}{2b}exp\left(-\frac{|x|}{b}\right)$.
\end{deff}

\begin{deff}[Gaussian Mechanism]
Given any function $f: \mathbb{N}^{|\mathcal{X}|} \rightarrow \mathbb{R}^{k}$, and $0< \epsilon < 1$, the Gaussian mechanism is defined as 
$$ \mathcal{M}_{G}\left(x,f,\epsilon,\delta\right)=f(x) + (Y_{1},...,Y_{k})$$
where $Y_{i}$ are i.i.d. random variables drawn from $\mathcal{N}\left(0,c\Delta_{2}(f)/\epsilon \right)$, with  $c^{2} > 2\ln(1.25/\delta).$
\end{deff}

\begin{remark}
Those mechanism transform a deterministic query $(f)$ into a randomized algorithm $(\mathcal{A})$ by adding random variables to the deterministic query.
\end{remark}

Given a privacy degree $(\epsilon,\delta)$, this two mechanisms respectively allow to release the answer to any numerical query while ensuring $\epsilon$-differential privacy and $(\epsilon,\delta)$-differential privacy. For a complete theoretical analysis of the two mechanisms we refer the reader to~\cite[Chapter~3]{Dwork_2013} and~\cite{Nissim_2007}.\\

The main issue while dealing with private algorithm design is the trade-off between privacy and accuracy. In fact, it is quite logical to think that the more accurate the queries' answers are, the more information one gets about the database, and therefore, the more privacy is degraded.

\begin{claim}[\cite{Dwork_2013} The Fundamental Law of Information Recovery] 
\label{Claim:TheFundamentalLawInformationRecovery}
"Overly accurate answers to too many questions will destroy privacy in a spectacular way" 
\end{claim}

This statement is a simple reformulation of the theoretical works about reconstructions attacks against private databases~\cite[chapter~8]{Dwork_2013} and~\cite{Dinur:2003:RIW:773153.773173}. It simply states that accuracy and privacy are not trivially combined.
To have a better understanding of what this trade-off is about, let us present the main result on the accuracy of the Laplace mechanism.

\begin{theorem}[\cite{Dwork_2006} Accuracy of the Laplace mechanism] 
Let $f:\mathbb{N}^{|\mathcal{X}|} \rightarrow \mathbb{R}^{k}$, $x \in \mathbb{N}^{|\mathcal{X}|}$, and $\epsilon >0$. Moreover let $y=\mathcal{M}_{L}(x,f,\epsilon)$. Then $\forall \gamma \in (0,1]$ : $$\mathbb{P} \left[ ||f(x) - y||_{\infty} \geq \frac{\Delta_{1}(f)}{\epsilon}\ln\frac{k}{\gamma} \right] \leq \gamma, \textit{ $k$ being the dimensionality of the output space.}$$
\end{theorem}

Seeing this theorem, it is quite straightforward that the accuracy is non-increasing regarding $\epsilon$. Therefore a strong privacy implies less accurate outcomes and vice versa. The accuracy/privacy tradeoff is a central issue that have to be tackled for any new differentially private method, and this will be an important concern for the introduction of our contribution in Chapter~\ref{section:PrivateMSTs}.
However, now that one is more familiar with the basic constructions let us introduce the main tool to construct complex differentially private mechanisms: the composition. \\

The main strength of differential privacy is its ability to compose, i.e to construct a complex differentially private mechanism from several simple ones. So far we presented a definition of differential privacy and two of the simplest ways to construct a differentially private mechanism, let us now introduce how to compose such simple blocks.

\begin{prop}[e.g~\cite{Dwork_2013} Post-Processing] 
\label{postprocessing}
Let $\mathcal{A}$ be a $(\epsilon,\delta)$-differentially private randomized algorithm with domain $\mathbb{N}^{|\mathcal{X}|}$  and range $B$, and $h : B \rightarrow B'$ a deterministic (measurable) mapping.
Then $h \circ \mathcal{A}$ is $(\epsilon,\delta)$-differentially private.
\end{prop}

This proposition is clearly a strong result that allows one to compute a one-time-only release of the private mechanism for any other deterministic algorithm to be performed. One can also compose several $(\epsilon,\delta)$-differentially private mechanisms according to the following propositions.

\begin{prop}[\cite{Dwork_2013} Simple Composition] 
Let $k \in \mathbb{N}$ and $(\mathcal{A}_{i})_{i\in [k]}$ such that for all $i \in [k], \mathcal{A}_i$ is an $(\epsilon_{i},\delta_{i})$-differentially private algorithm with domain $\mathbb{N}^{|\mathcal{X}|}$, and range $B_{i}$. Then $\mathcal{A}_{[k]}$  is an $\left(\sum_{i=1}^{k}\epsilon_{i},\sum_{i=1}^{k}\delta_{i} \right)$-differentially private algorithm with range $\prod\limits_{ i\in [k]} B_{i}$, where $ \forall x \in \mathbb{N}^{|\mathcal{X}|}$, $\mathcal{A}_{[k]}(x)=\left(\mathcal{A}_{i}(x) \right)_{i\in [k]}$.
\end{prop}
\begin{prop}[\cite{dwork2006our} Adaptive Composition] 
\label{basiccomposition}
For any $\epsilon, \delta \geq 0$ the adaptive composition of $k$ $(\epsilon,\delta)$-differentially private mechanisms is $(k\epsilon,k\delta)$-differentially private.
\end{prop}

Once more, one must recall that composition is easy but degrade either the accuracy or the privacy of the resulting mechanism according to Claim~\ref{Claim:TheFundamentalLawInformationRecovery}, it should thus be used cautiously. The main keys for differentially private analysis on numerical databases being set, let us now introduce the basics of the differentially private analysis on structured databases.

\section{The Exponential mechanism}

First introduced by McSherry and Talwar in~\cite{mechanism-design-via-differential-privacy}, the Exponential mechanism is the typical way of answering arbitrary range queries while preserving differential privacy. It was designed for the situations were simply adding noise to the computed quantity is rather impossible (non-numeric data) or would destroy the complete data structure (in game theory slightly modifying the strategy can make the outcome drop from optimal to null~\cite{fudenberg1991game}).
Given some range $\mathcal{R}$ of possible responses to a query, it is defined according to a utility function $u:= \mathbb{N}^{\mathcal{|X|}} \times \mathcal{R} \rightarrow \mathbb{R}$. Intuitively, it does not need to be a utility function in terms of game theory, but it aims at providing some total preorder on the range $\mathcal{R}$ according to the total order of $\mathbb{R}$. The sensitivity of this function is denoted $$\Delta u = \max\limits_{r \in \mathcal{R}} \max\limits_{ x \sim y \in \mathbb{N}^{\mathcal{|X|}}} | u(x,r) - u(y,r)|.$$

\begin{deff}[Exponential mechanism]
Given some output range $\mathcal{R} $, a privacy parameter $\epsilon > 0$, an utility function $u:= \mathbb{N^{\mathcal{|X|}}} \times \mathcal{R} \rightarrow \mathbb{R}$, and some $x \in \mathbb{N}^{\mathcal{|X|}} $, the exponential mechanism $\mathcal{M}_{Exp}(x, u, \mathcal{R}, \epsilon)$ selects and outputs an element $r \in \mathcal{R}$ with probability proportional to $\exp\left(\frac{\epsilon u(x,r)}{2 \Delta u}\right)$.
\end{deff}

The exponential mechanism defines a distribution on a potentially complex and large range $\mathcal{R}$. As the following proposition states,  sampling from such a distribution preserves $\epsilon$-differential privacy. 

\begin{prop}[\cite{mechanism-design-via-differential-privacy} Privacy of the Exponential mechanism]
\label{prop:exponentialprivacy}
For any non empty range $\mathcal{R}$, the exponential mechanism preserves $\epsilon$-differential privacy, i.e. if $x\sim y \in \mathbb{N^{\mathcal{|X|}}}, $ $$\mathbb{P}\left[\mathcal{M}_{Exp}(x, u, \mathcal{R}, \epsilon) =r\right] \leq e^{\epsilon}\mathbb{P}\left[\mathcal{M}_{Exp}(y, u, \mathcal{R}, \epsilon) =r\right]. $$
\end{prop}

As stated in section~\ref{section:ClassicalResults},  every private mechanism has to achieve a trade-off between privacy and accuracy of the method. The exponential mechanism does not constitute an exception to this rule. The following proposition highlights this trade-off for the exponential mechanism when $|\mathcal{R}| <  +\infty$.

\begin{prop}[reformulation~\cite{Dwork_2013} Accuracy of the Exponential mechanism]
\label{tradeoffexponential}
Given some dataset $x \in \mathbb{N}^{\mathcal{|X|}}$, a non empty output range $\mathcal{R}$, a privacy parameter $\epsilon > 0$, a utility function $u:= \mathbb{N}^{\mathcal{|X|}} \times \mathcal{R} \rightarrow \mathbb{R}$, and denoting $OPT_{u}(x)=\max\limits_{r \in \mathcal{R}}u(x,r)$, then for all $t \in \mathbb{R}$
$$\mathbb{P}\left[u(x,\mathcal{M}_{Exp}(x,u,\mathcal{R},\epsilon))\leq OPT_{u}(x) - \frac{2\Delta u}{\epsilon}\left(\ln\left(|\mathcal{R}|\right) + t\right)\right] \leq e^{-t}. $$
\end{prop}

The exponential mechanism opens up the field of possibilities about what differential privacy on structured databases could be. Chapter~\ref{section:DPforGraphs} focuses on investigating differential privacy on graph databases, and presents the privacy type we have chosen to introduce in the remaining of this work.
 
\chapter{Differential privacy for graph databases}
\label{section:DPforGraphs}

The neighborhood notion between two databases is the central notion of the definition of differential privacy, therefore, as soon as one gives a notion of neighboring between two (structured or unstructured) databases, one can construct differentially private mechanisms to protect the sensitive information within those databases. Accordingly, how to consider differential privacy on structured databases has become a fundamental question, due to the diversity and complexity of the attributes one could want to keep private. Mir and al. in~\cite{5360515} as well as Karwa and al. in ~\cite{karwa2011private} formalized the idea of releasing statistics on a graph in a differentially private manner following the seminal work of Nissim and al. in~\cite{Nissim_2007}.
Furthermore the community focused on how to sanitize the graph, therefore several definitions of neighboring and privacy on graphs appeared. Chapter~\ref{section:DPforGraphs} focuses on introducing the different definitions of differential privacy for graph stressing the presentation of weight-differential privacy which will be used in the remaining of this work.

\section{Edge and node differential privacy}
\begin{figure}[!ht]
\centering
\includegraphics[width=\textwidth]{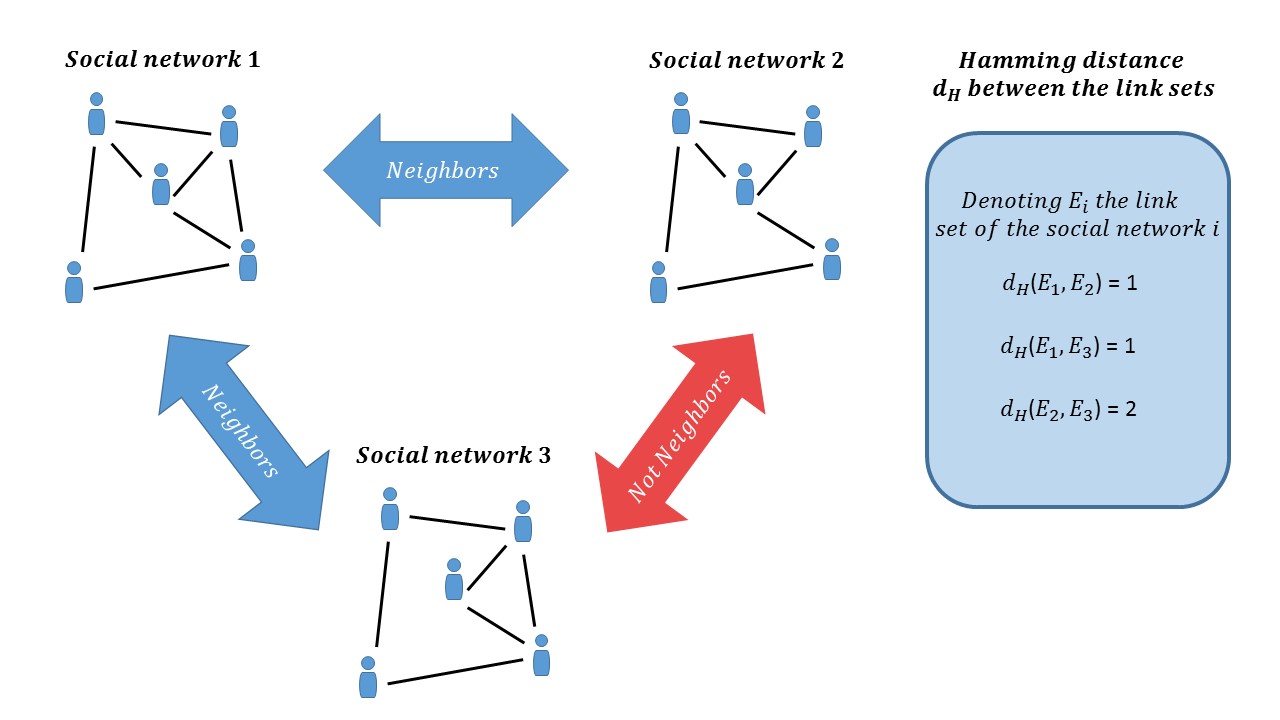}
\caption{Evaluation of the neighboring between social networks based on the Hamming distance between the link sets.}
\label{figure:edgeDP}
\end{figure}

The two common ways of conceiving privacy on a graph topology are 
edge-differential privacy~\cite{5360242}, and node-differential privacy~\cite{Kasiviswanathan2013}. Both are strongly connected to the analysis of social networks. Under edge-differential privacy, two graph topologies are neighbors if one of them can be obtained from the other by adding or removing a single edge. 
This kind of neighborhood notion, and the privacy definition that follows could formalize the hiding of a friendship relation between to individuals as illustrated by Figure~\ref{figure:edgeDP}.
This figure also presents the link between the intuition and the formal definition of edge-neighboring graph topologies based on the Hamming distance between the edge sets.

\begin{deff}[Edge-neighboring graph topologies]
Two graph topologies $\mathcal{G}_{1}=(V_{1},E_{1})$ and $\mathcal{G}_{2}=(V_{2},E_{2})$, are said to be edge-neighboring, denoted $\mathcal{G}_{1}\underset{edge}{\sim} \mathcal{G}_{2}$ if $$ \min\limits_{\sigma \in Perm(E) }d_{H}\left[\sigma(E_{1}),\sigma(E_{2})\right] \leq 1, $$ i.e if the minimum Hamming distance between any two permutation $\sigma$ of their edge sets is lower or equal to one.
\end{deff}

\begin{remark}
For simplicity one could denote $d_{H}(E_{1},E_{2})=\min\limits_{\sigma \in Perm(E) }d_{H}\left[\sigma(E_{1}),\sigma(E_{2})\right]$ making the assumption that the edge sets are "well ordered" (as in Figure~\ref{figure:edgeDP}).
\end{remark}

\begin{figure}[!ht]
\centering
\includegraphics[width=\textwidth]{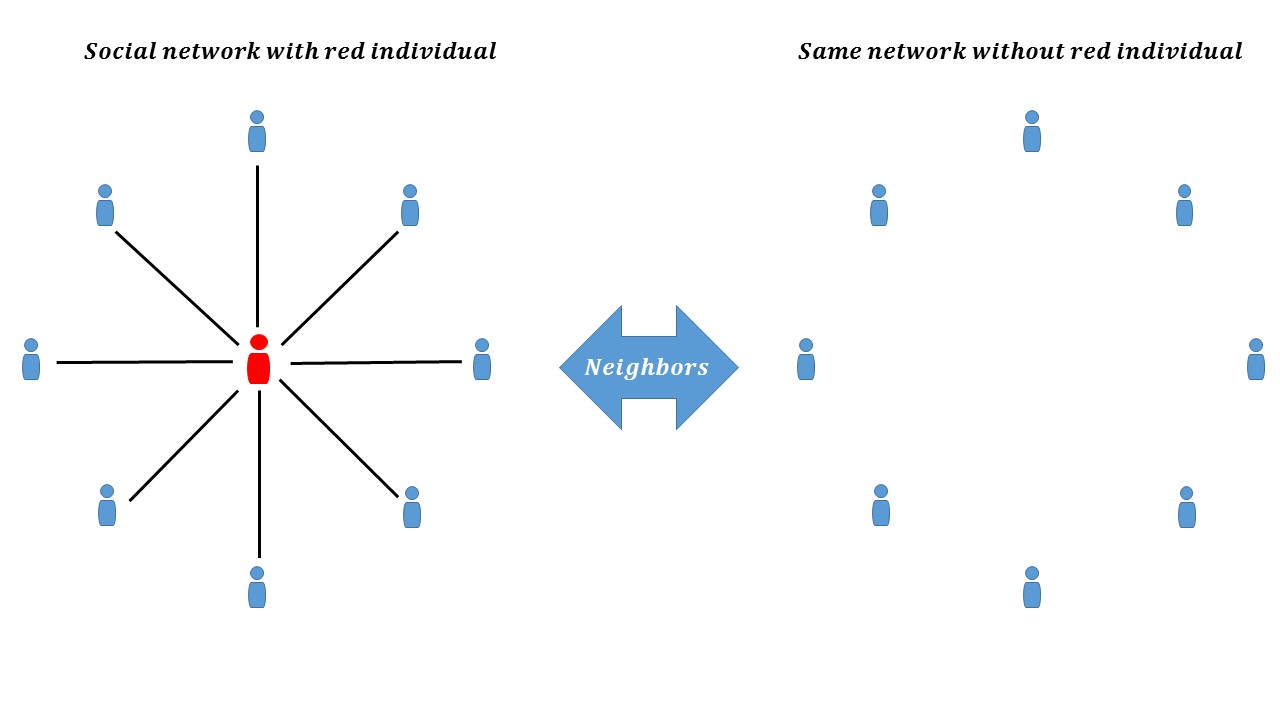}
\caption{Two node-neighboring social networks with strongly differing topologies}
\label{figure:nodeDP}
\end{figure}

For node-differential privacy, two graphs are said to be neighbors if one can be obtained from the other by removing a single node and all its incident edges. This could formalize the presence of an individual in a social network as Figure~\ref{figure:nodeDP} illustrates.

This neighborhood notion is a more permissive than the previous one, since it allows two highly different topologies to be neighboring as illustrated in Figure~\ref{figure:nodeDP}. The formal definition of node differential privacy is also constructed according to the Hamming distance, but between the node sets. 
Although edge and node differential privacies are widely used on structured databases, both of them are close to social network analysis, and thus focused on the graph topologies. 
The privacy choices are mainly made according to the nature of the data at hand, and many real life situations such as flow monitoring in energy-supply networks or transport networks analysis can be represented by weighted graphs with fixed topology and evolving weights.
Section~\ref{section:WeightDP} presents a new definition of privacy for such structured databases where edge or node privacy are not applicable.

\section{Weight-differential privacy}
\label{section:WeightDP}
The idea of weight-differential privacy seems to first appear in an implementation paper by Costea and al.~\cite{6511749} , but theoretical foundations for considering weight-differential privacy on a graph has been recently introduced by Adam Sealfon in~\cite{Sealfon_2016}.
The emergence of this new notion is related to the existence of some settings (e.g transport networks) in which the structure of the database is intrinsic to the problem but fixed, and a lot of the information is carried by the edge weights. In such settings the privacy breach will occur on the edge weights, therefore edge or node-differential privacy are not applicable, and this is how weight-differential privacy emerged. Example~\ref{example:weightDP} aims at providing a good intuition of the settings where the information carried by the weights has to be protected.

\begin{exemple}
\label{example:weightDP}
An analyst wants to produce a study of the traffic within a subgraph of the World Wide Web. In this setting, the flow between two websites is the number of users passing from one to the other using an existing hyper-link.
A simplified version of this problem can be formalized by the simple-undirected-weighted graph $G=(V,E,w)$, where all nodes are websites and the edges are the existing hyper-links between them. The weight function $w:= E \rightarrow \mathbb{R}^{+}$ returns for any edge, given a chosen population, the number of individuals passing, one way or the other, through the corresponding hyper-link in the world wide web network. 
 
  \begin{figure}[ht]
    \includegraphics[width=\textwidth]{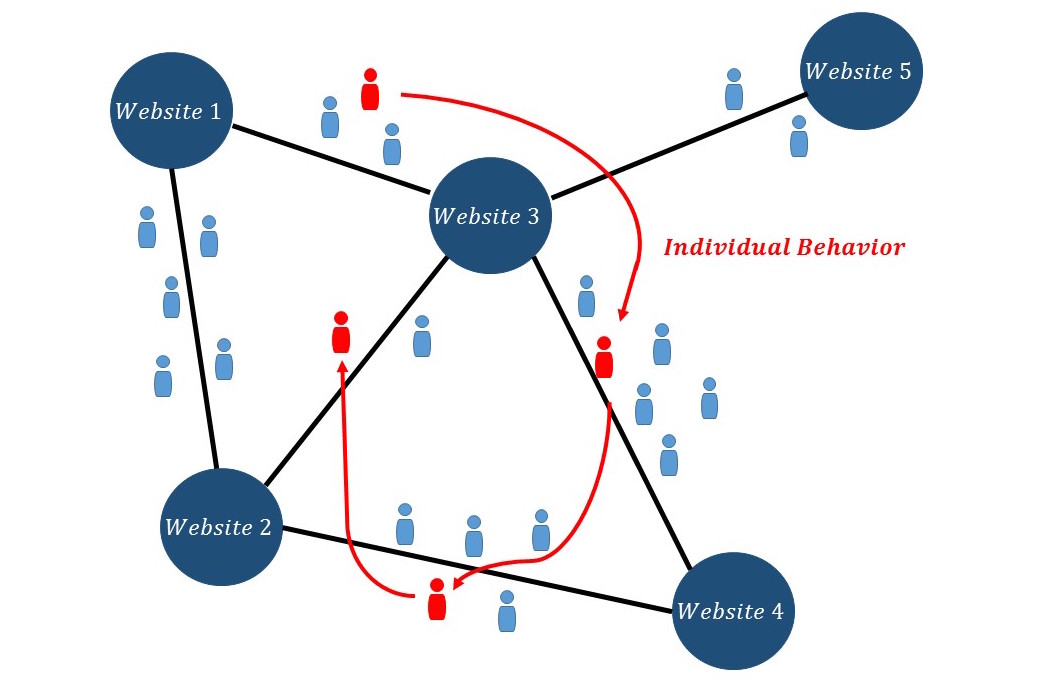}
    \caption{A subgraph of the World Wide Web weighted by the individual passing via an hyper-link from one site to the other. Red path highlights one individual behavior.}
     \label{figure:subgraphWWW}
    \end{figure}
  
  \textbf{What kind of privacy breach could occur?} For the purpose of the example, let us suppose that in Figure~\ref{figure:subgraphWWW},  \textit{Website 3} is a high sensitive website (political, or sexual orientation related for example). The analyst needs to have some statistics about this website attendance, for example the value of the incident hyper-link on which the traffic is the highest, but he is not supposed to be able to indicate whether some particular individual is related to this website or not, given the sensitive aspect of such an information. However, in absence of a curator applying some mechanism ensuring privacy, if he chooses to produce a differencing attack on  the red individual in the graph given the statistic "what is the value of highest weighted incident edge" he will infer the personal behavior of the red individual concerning \textit{Website 3}. 
  
\end{exemple}

Motivated by similar settings, Adam Sealfon gives the formal definition of weight-differential privacy in~\cite{Sealfon_2016}. He has also introduced algorithms and theoretical bounds for the release of distances, paths, matchings, and spanning trees under privacy constraints. The rest of Chapter~\ref{section:DPforGraphs} is dedicated to presenting the main definitions and the privacy mechanism he introduced, whereas Chapter~\ref{section:PrivateMSTs} presents, among other things, his result on the release of an approximated minimum spanning tree.\\

Let $G=(V,E,w)$ be a weighted graph with a vertex set $
V$, an edge set $E$, and a weight function $w:= E \rightarrow \mathbb{R}$ mapping any edge in $E$ to its weight in the graph. Since weight-differential privacy is not a topological based privacy definition, the neighborhood notion is mostly related to the weights.

\begin{deff}[$\boldsymbol{\ell_{1}}$-neighboring between weight functions]
\label{definition:Selfonweightneighboring}
For any edge set $E$, two weight functions $ w,w' \in \mathcal{W}_{E}$ are neighboring, denoted $ w \sim w'$, if  $$|| w - w' ||_{1}:=\sum\limits_{e \in E}| w(e) - w'(e) | \leq 1.$$ 
\end{deff}

Section~\ref{section:NewPrivacy} presents our contribution and reformulation of this definition, this is why, for clarity of the remaining of this work we call this neighboring $\ell_{1}$-neighboring, denoted $ w \underset{\ell_{1}}{\sim} w'$.

\begin{deff}[$\boldsymbol{\ell_{1}}$-weight-neighboring graphs]Let $G=(V,E,w)$ and $G'=(V',E',w')$, two weighted graphs, $G$ and $G'$ are said to be $\ell_{1}$-weight neighboring if $V=V'$, $E=E'$ and  $ w \underset{\ell_{1}}{\sim} w'$.
\end{deff}

Now, let one formally define weight-differential privacy for graph algorithms according to the above notions.

\begin{deff}[$\boldsymbol{\ell_{1}}$-weight-differential privacy]
For any graph topology $\mathcal{G}= (V,E)$, let $\mathcal{A}$ be a randomized algorithm that takes as input a weight function  $w\in \mathcal{W}_{E} $. $\mathcal{A}$ is called $(\epsilon,\delta)$-$\ell_{1}$-weight-differentially private on $\mathcal{G}= (V,E)$ if for all pairs of $\ell_{1}$-neighboring weight functions $ w,w'\in \mathcal{W}_{E} $, and for all set of possible outputs S, one has $$\mathbb{P}\left[\mathcal{A}(w) \in S\right] \leq e^{\epsilon}\mathbb{P}\left[\mathcal{A}(w') \in S\right] + \delta $$

If $\mathcal{A}$ is $(\epsilon,\delta)$-differentially private on every graph topology in a class $\mathcal{C}$, it can be denoted as $(\epsilon,\delta)$-differentially private on $\mathcal{C}$.
\end{deff}

Weight-differential privacy is a very new research field. For now the privacy of most graph statistics~\cite{Sealfon_2016,brunetnovel,brunet2016edge} are based on a sanitizing of the graph using a pre-processing. This means that a generic mechanism is used to ensure weight-privacy of all the graph weights before using a post-processing (Proposition~\ref{postprocessing}) to compute any other statistic. Those generic mechanisms are immediate applications of the Gaussian or Laplace mechanism as defined in Section~\ref{section:ClassicalResults}. More precisely weight-differential privacy is ensured by using the simple idea of adding identically distributed random noise drawn from a Laplace/Gaussian distribution on every edge weight. 
 
\begin{deff}[Graph-based Laplace mechanism]
Let $G=(V,E,w)$ be a weighted graph, and $\epsilon > 0$ a privacy parameter. Graph-based Laplace mechanism is defined as
$$ \mathcal{M}_{GbL}(G,\epsilon)=G'=(V,E,w'=w+ (Y_{1},...,Y_{|E|}))$$
where $Y_{i}$ are i.i.d. random variables drawn from $Lap\left(1/\epsilon \right)$. 
With $w+ (Y_{1},...,Y_{|E|})$ meaning that if one gives an arbitrary order to the edges $E=(e_{i})_{i \in [|E|]}$, one has $w'(e_{i})=w(e_{i}) + Y_{i}$ $\forall  i \in [|E|]$.
\end{deff}

\begin{remark}
This graph based mechanism is a direct application of the Laplace mechanism from Section~\ref{section:ClassicalResults} considering that the numerical query $f$ is the releasing of all the arbitrarily ordered weights. Therefore, the method inherits all the properties from Section~\ref{section:ClassicalResults} about the privacy and accuracy of the Laplace mechanism.
\end{remark}

\begin{remark}
The analog analysis is available for the Gaussian mechanism in~\cite{brunetnovel,brunet2016edge} with a fine analysis of the block-structured sensitivity of the graph.
\end{remark}

The generic mechanisms allow one to release a noisy version of the initial graph, which will make any post-processing $\epsilon$-differentially private. They are useful, but since the information release is massive (every graph weight is released with the graph structure), the total amount of noise needed to keep the graph private is huge. As claimed in Section~\ref{section:ClassicalResults}, such a generic sanitizing mechanism cannot ensure close theoretical and experimental bound on every statistics an analyst could need. Chapter~\ref{section:PrivateMSTs} presents, based on the minimum spanning tree statistic, how one can choose to privately construct a selected statistic instead of sanitizing the whole graph and computing this statistic as a post processing.

\chapter{Private Minimum Spanning Tree}
\label{section:PrivateMSTs}

In this chapter, we firstly present the main result on releasing the minimum spanning tree in a weighted graph under privacy constraints, and secondly introduces a new algorithm to tackle the issue of privately releasing an approximated minimum spanning tree topology for a simple-undirected-weighted graph. Stressing the theoretical and experimental comparison between the Post-processing method (Laplace mechanism) and the In-processing one (new algorithm) highlights the fact that In-processing, while being less generic, obtains better results and ensures the same privacy guarantees for this specific task. 

\section{In-processing versus Post-processing }
\label{section:InprocessingVSPostprocessing}
According to the framework of differential privacy, the analyst only has a query access to the database. The main difference between the In-processing access and the Post-processing one is the nature of the query. 

\begin{figure}[ht]
    \includegraphics[width=\textwidth]{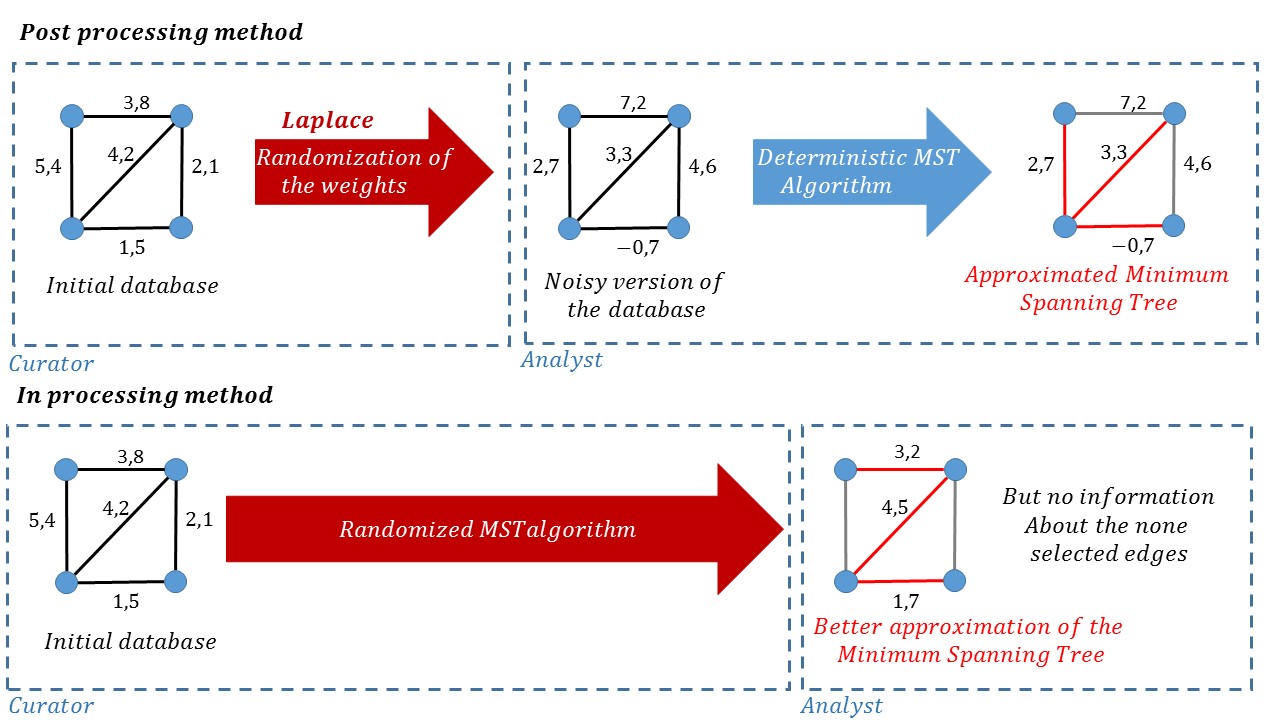}
    \caption{Comparison of in processing and post processing methods for the computation of a private minimum spanning tree (MST).}
     \label{figure:InprocessingVsPostprocessing}
\end{figure}

As illustrated by Figure~\ref{figure:InprocessingVsPostprocessing}, in a Post-processing method, the analyst asks one time for all the weights, and obtains a noisy version of them. Then he/she produces any deterministic treatment he/she likes (here a minimum spanning tree) given that the privacy is guaranteed on the weight release itself (Proposition~\ref{postprocessing}). In the In-processing method, the analyst directly asks the curator to produce a private version of the statistic he/she wants, and thus allows the curator to adapt the randomized construction to ensure the same privacy with better accuracy. The intuitive idea that supports the In-processing methods is that releasing a synthetic version of the database (here a minimum spanning tree) contains less information than releasing the whole database and thus sanitizing it is less costful in terms of accuracy.
The remaining of this chapter presents a Post-processing and an In-processing method for the releasing of a minimum spanning tree topology and compare theoretically and experimentally their accuracies.

\section{Approximated Minimum Spanning Tree by Graph-based Laplace mechanism}
While Nissim, Rakhodnika and Smith~\cite{Nissim_2007} only managed to privately compute the approximated cost of a minimum spanning tree, Sealfon~\cite{Sealfon_2016}, by using the Graph-based Laplace mechanism,  manages to release a true weighted tree under $\epsilon$-weight-differential privacy. His method is quite simple and amounts to use the Laplace mechanism to sanitize all the graph weights before computing a deterministic MST algorithm on the new graph. He claims the topology of the minimum spanning tree obtained on the sanitized graph to be almost minimal weight if it is evaluated with the real weights (non sanitized weights). To evaluate the accuracy of his method, he compares the weight of a minimum spanning tree in the initial graph (before Laplace mechanism) with the weight (using the non sanitized weight function) of the tree-topology that minimizes the total noisy-weight (the topology of the MST in the sanitized graph).
Definition~\ref{definition:error} should clarify and formalize this error evaluation.

\begin{deff}[Approximation Error for Approximated Minimum Spanning Tree]
\label{definition:error}
Let $G=(V,E,w)$ be a weighted graph, $\epsilon>0$, $\mathcal{T}$ the topology of the minimum spanning tree of $G$ and $\mathcal{T'}$ the topology of an approximated minimum spanning tree obtained by an arbitrary randomized algorithm. Since $\mathcal{T}$ and $\mathcal{T'}$ can be characterized by two edge subsets of $E$ : $\mathcal{T}=(e_{i})_{i \in [ |V|-1 ]}$ and $\mathcal{T'}=(e'_{i})_{i \in [ |V|-1 ]}$, the error of the spanning tree topology $\mathcal{T'}$ for the approximation of the minimum spanning tree topology of $G$ is $$Error_{G}\left(\mathcal{T'}\right)=\sum\limits_{i \in [ |V|-1 ]}w(e'_{i}) - \sum\limits_{i \in [ |V|-1 ]}w(e_{i}).$$
\end{deff}

\begin{remark}
For simplicity one can write the total weight of a spanning tree $\mathcal{T}$ as $w(\mathcal{T})$.
\end{remark}
Using Definition~\ref{definition:error}, we reformulate, for the sake of clarity, in the following theorem, the theoretical worst case guarantee on the error of an approximated minimum spanning tree topology produced by Sealfon's method~\cite{Sealfon_2016} (Graph-based Laplace mechanism + deterministic MST algorithm).

\begin{theorem}[\cite{Sealfon_2016} Error of a Graph-based Laplace mechanism  AMST]
\label{theorem:Sealfonbound}
 For any $\epsilon >0, \gamma \in (0,1]$, $G=(V,E,w)$, and $\mathcal{T'}$ the topology of a minimum spanning tree on $G'=\mathcal{M}_{GbL}(G,\epsilon)$, one has $$\mathbb{P}\left[Error_{G}\left(\mathcal{T'}\right) \geq \frac{2(|V|-1)}{\epsilon}\ln\frac{|E|}{\gamma} \right] \leq \gamma.$$
\end{theorem}

This theorem evaluates the accuracy of the topology of the minimum spanning tree on $G'$ given the weight of the same topology in $G$. Since this error definition does not depend on the mechanism used to produce the approximated minimum spanning tree topology, we also use it to evaluate the error the In-processing algorithm makes. The following section details our contribution to the theory of weight-differential privacy, compares the theoretical bound obtained with the bound in Theorem~\ref{theorem:Sealfonbound}, and produces some experiments to support this comparison.

\section{A new privacy setting}
\label{section:NewPrivacy}

The model for privacy presented in this section is similar to the one in~\cite{Sealfon_2016}, given that the same definition of differential privacy is used. Nevertheless the neighborhood definition used to compare two graphs relies on the $\ell_{\infty}$ distance between the weight function outputs, while the previous model imposed the whole graph changes to sum to 1 inducing a neighborhood notion based on the $\ell_{1}$ distance.
The mechanism and definitions presented are a straightforward applications of the preliminaries from Chapter~\ref{section:IntroToDP}, they thus inherit all the proprieties introduced in this chapter.
First, to fully understand the need to introduce such a new definition, one must present the framework from which the idea of a new definition emerged. \\

\begin{figure}[!ht]
\centering
\includegraphics[width=\textwidth]{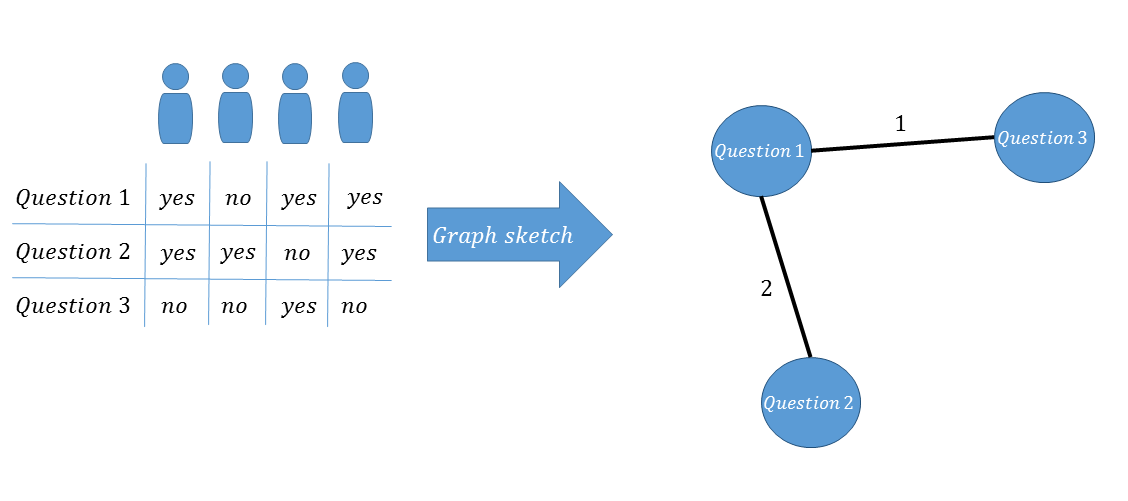}
\caption{Graph sketching framework applied on a population of size 4 (N) and 3 questions (M). Only the co-occurrence of the response "yes" between the question 1 and the other questions is observed (E)}
\label{figure:graphsketch}
\end{figure}

\hspace{-1cm}
\fbox{
\begin{minipage}[b]{1.01\linewidth}
\vspace{0.2cm}
\textbf{Graph sketching Framework:} 
Let us consider the new general framework that could be the formal vision of Example~\ref{example:weightDP} as well as individuals answering a survey by yes or no questions as illustrated by Figure~\ref{figure:graphsketch}  \\

One has a dataset of vectors $(z_{i})_{i \in [N]}$, where $z_{i} \in \{0,1\}^{M}$ (we denote the $j^{th}$ element  of the vector $z_{i}$ by $z_{i}^{j}$).
One does not need to assume anything about the size of $N$ and $M$ else than $N$ > $M$.
To efficiently represent this database for a correlation analysis, one could want to sketch it into a graph. Such a graph could be as follows: \\$G=(V,E,w)$ with $V=\{1,...M\}$ (one node per variable), $E$ is the set of couples of variables for which the analyst want to observe the correlation, and finally the weight function is characterized by $ w\left(\left(i,j\right)\right)= \#\{k \in [N] \textit{ s.t. } z_{k}^{j}=z_{k}^{i}=1 \}$ for every $(i,j) \in E$
(here we  consider counting edges but one can choose to adapt those weights to the partial correlation or any other similarity, it might change the sensitivity of the method but our method can scale up according to the sensitivity of the framework).\\
\end{minipage}
} \vspace{0.1cm}

In such a setting, the sensitive information is carried by every individual $z_{i}$, therefore, to protect the individuals, privacy should be respected on the edges' weights. Nevertheless, using the definition from~\cite{Sealfon_2016} would be too costful in terms of accuracy since it adds noise scaled to $M$ to every edges (in fact, the change of one individual in the database could change the $\ell_1$ distance between weight function by at most $M$), potentially destroying the whole weight structure of the graph. In such settings, we give a new privacy definition that will allow us to produce a strongly accurate almost minimum spanning tree topology.

\begin{deff}[$\boldsymbol{\ell_{\infty}}$-Neighboring weight functions] For any edge set $E$, two weight functions $w,w'\in \mathcal{W}_{E} $ are $\ell_{\infty}$-neighboring, denoted $ w \underset{\ell_{\infty}}{\sim} w'$, if  $$ || w - w' ||_{\infty} := \max\limits_{e \in E}| w(e) - w'(e) | \leq 1/2.$$  \end{deff}

The bound is set to $1/2$ for simplicity of the proofs, it also allows one to consider an  $\ell_{\infty}$ ball of diameter $1$ defined around the real weight function by analogy to the $\ell_{1}$ ball from Definition~\ref{definition:Selfonweightneighboring}. Moreover, as will be explained later, the results we obtain are scalable since they trivially adapt to the initial sensitivity of the weights, therefore this arbitrary choice of bound is not an issue. The neighborhood between graphs and the definition of $\ell_{\infty}$-weight-differential privacy directly adapts from those introduced in Section~\ref{section:WeightDP}.

\begin{deff}[$\boldsymbol{\ell_{\infty}}$-Weight differential privacy]
For any graph topology $\mathcal{G}= (V,E)$, let $\mathcal{A}$ be any randomized algorithm that takes as input a weight function  $w\in \mathcal{W}_{E} $. $\mathcal{A}$ is called $(\epsilon,\delta)$-$\ell_{\infty}$-weight differentially private on $\mathcal{G}= (V,E)$ if for all pairs of $\ell_{\infty}$-neighboring weight functions $ w,w'\in \mathcal{W}_{E} $, and for all set of possible outputs $S$, one has $$\mathbb{P}\big[\mathcal{A}(w) \in S\big] \leq e^{\epsilon}\mathbb{P}\big[\mathcal{A}(w') \in S\big] + \delta.$$

If $\mathcal{A}$ is $(\epsilon,\delta)$-differentially private on every graph topology in a class $\mathcal{C}$, it can be denoted as $(\epsilon,\delta)$-differentially private on $\mathcal{C}$.
\end{deff}

Finally one should introduce the graph-based exponential mechanism. For the remaining of this document, the graph-based exponential mechanism is referred to as exponential mechanism specifying only if it is necessary.

\begin{deff}[Graph-based Exponential mechanism]
Given some graph topology $\mathcal{G}=(V,E)$, an output range $\mathcal{R} \subset E$, a privacy parameter $\epsilon > 0$, an utility function $u_{\mathcal{G}}:= \mathcal{W}_{E} \times \mathcal{R} \rightarrow \mathbb{R}$, and some $w\in \mathcal{W}_{E}$ the graph-based exponential mechanism $\mathcal{M}_{Exp}(\mathcal{G}, w, u_{\mathcal{G}}, \mathcal{R}, \epsilon)$ selects and outputs an element $r \in \mathcal{R}$ with probability proportional to $exp\left(\frac{\epsilon u_{\mathcal{G}}(w,r)}{2 \Delta u}\right)$.
\end{deff}

The Exponential mechanism and the new definitions of privacy on weighted graphs are the key elements of our new algorithm for producing an almost minimum spanning tree topology under differential privacy constraints. We call the algorithm \textsc{PAMST}.

\section{\textsc{PAMST} algorithm}
\textsc{PAMST} relies on a Prim-like minimum spanning tree construction ensuring that  at every iteration of the algorithm the nearest neighbor is not automatically chosen, but the exponential mechanism provides an approximated nearest neighbor under differential privacy constraints. This algorithm implies an iterative use of the exponential mechanism, it is thus necessary to define what utility function this mechanism is going to use.

Let $G=(V,E,w)$ be a simple-undirected-weighted graph, $\mathcal{G}$ its topology, and $\mathcal{R}$
a set of edges from $E$. The utility function used in the iterative call of the exponential mechanism is 
$$ \begin{array}{ccccc}
u_{\mathcal{G}} & : & \mathcal{W}_{E} \times \mathcal{R} & \to & \mathbb{R} \\
& & (w,r) & \mapsto & - |w(r) - \min\limits_{r' \in \mathcal{R}} w(r')|  \\
					\end{array}$$
Given the above setting, without normalization on the edges weight, one has $\Delta u_{\mathcal{G}} = 1$ for any non empty $\mathcal{R}$ (the change of an individual in the database can change each weight by at most $1$), and with normalization $\Delta u_{\mathcal{G}} = 1/|E|$. For the sake of comparison to the Laplace Post-processing~\cite{Sealfon_2016}, normalization is needed to obtain a maximum global variation of the weights less or equal to 1 (in such a setting one can consider $\ell_{\infty}$ differential privacy as a particular case of $\ell_{1}$ differential privacy).
The intuition of why it is possible to release an exact-weight-based spanning tree is as follows: the exponential mechanism, by using an utility function to give some pre-order on $\mathcal{R}$ transfers the private information contained in $\mathcal{R}$ to an order in $\mathbb{R}$. Thus, to keep the mechanism differentially private, it suffices to keep the outputs of $u$ private instead of the values from $\mathcal{R}$.\\

\textbf{Additional notations:} Let $S$ be a set of nodes from $G$, and $\mathcal{R}_{S}$ the set of edges that is incident to one and only one node in $S$ (also denoted xor-incident). For any edge $r$ in such a set, the incident node to $r$ that is not in $S$ is denoted $r_{\rightarrow}$. T
Finally, the restriction of the weight function to an edge set $\mathcal{R}$ is denoted $w_{|\mathcal{R}}$.

\begin{algorithm}[H]
\caption{Private Approximated Minimum Spanning Tree \textsc{PAMST}$(\mathcal{G},u_{\mathcal{G}},w,\epsilon)$}
\label{alg:Private-MST}
\begin{algorithmic} 
\REQUIRE A graph topology $\mathcal{G}=(V,E)$, a weight function $w$, a degree of privacy $\epsilon$, utility function $u_{\mathcal{G}}$. 
\ENSURE The topology of an approximated minimum spanning tree represented by $S_{E}$ a set of edges.

\STATE 
\STATE Pick $v \in V$ arbitrarily 
\STATE $S_{V} \leftarrow \{v\}$
\STATE $S_{E} \leftarrow \emptyset$
\WHILE{$S_{V}\neq V$ }
\STATE $r=\mathcal{M}_{Exp}(\mathcal{G},w,u_{\mathcal{G}},\mathcal{R}_{S_{V}},\frac{\epsilon}{|V|-1})$
\STATE $S_{V} \leftarrow S_{V} \cup \{r_{\rightarrow}\}$
\STATE $S_{E} \leftarrow S_{E} \cup \{r\}$
\ENDWHILE
\RETURN $S_{E}$
\STATE 
\end{algorithmic}
\end{algorithm}

\textsc{PAMST} takes as an input a graph topology, and a weight function associated to its set of edges. It outputs the topology of a spanning tree which weight according to the weight function is almost minimal.
To do so, the algorithm starts at an arbitrarily chosen vertex, and chooses one of its xor-incident edges according to the exponential mechanism and updates the set of edges with it. Then it adds the second node forming this edge to the current node set. While the node set does not contain all the nodes of the topology, PAMST continues to choose at each step an edge that is xor-incident to the current node set to update the edge set, and updates the node set accordingly.

Let us now introduce some results about the privacy and accuracy of \textsc{PAMST}. The privacy guarantees mostly rely on the properties of the exponential mechanism and the ability of differential privacy to compose, while the privacy/accuracy trade-off is less elementary.
The following theorems firstly presents the privacy guarantees, and secondly the accuracy of this new method in the $\ell_{\infty}$-weight differential privacy framework.

\begin{theorem}
Let $\mathcal{G}=(V,E)$ be the topology of a simple-undirected graph, then $\forall \epsilon >0$, $PAMST(\mathcal{G},\LargerCdot,\epsilon)$  is $\epsilon$-differentially private on $\mathcal{G}$  for $\ell_{\infty}$ neighborhood notions.
\end{theorem}

\begin{preuve}
Let $\mathcal{G}=(V,E)$ be a simple undirected weighted graph topology, $\epsilon$ a privacy degree, and $w \in \mathcal{W}_{E}$ a weight function. At every step of \textsc{PAMST}$\left(\mathcal{G},u_{\mathcal{G}},w,\epsilon \right)$,  $\mathcal{M}_{Exp}\left(\mathcal{G},w,u_{\mathcal{G}},\mathcal{R}_{S_{V}},\frac{\epsilon} {|V|-1}\right)$ is $\frac{\epsilon}{|V|-1}$-$\ell_{\infty}$-weight-differentially private by property of the Exponential mechanism (Proposition~\ref{prop:exponentialprivacy}). This property is independent of the neighborhood notion since the Exponential mechanism scales its distribution with $\Delta u$ which directly depends on the neighborhood notion at hand, here the $\ell_{\infty}$ neighboring.

Since Algorithm~\ref{alg:Private-MST} is no more than $|V|-1$ adaptive calls of $\frac{\epsilon}{|V|-1}$ differentially private mechanisms, Proposition~\ref{basiccomposition} gives us that Algorithm~\ref{alg:Private-MST} is $\epsilon$-differentially private on $\mathcal{G}$ regardless of the chosen neighborhood notion. \hfill\ensuremath{\blacksquare}
\end{preuve}

\begin{remark}
By notation abuse, one can state that this algorithm is $\epsilon$-weight-differentially private since it does not depend on the neighboring definition. In fact, even though one only focuses here on the $\ell_{\infty}$ setting, \textsc{PAMST} could also be $\epsilon$-$\ell_{1}$-weight-differentially private if used with an other framework. 
\end{remark}

Let us now introduce the theoretical bound for Algorithm~\ref{alg:Private-MST} that highlights the privacy/accuracy trade-off for this new method.

\begin{theorem}
\label{firstbound}
Let $G=(\mathcal{G},w)$, $w \in \mathcal{W}_{E}$, and $\epsilon > 0$. Let one denote $\{\mathcal{R}_{1},..., \mathcal{R}_{|V|-1}\}$ the ranges used in the successive calls of the exponential mechanism in \textsc{PAMST}$(\mathcal{G},u_{\mathcal{G}},w,\epsilon)$, $\mathcal{T}$ the spanning tree topology returned by this algorithm and $w(\mathcal{T})$ its associated total weight, then one has
$$w(\mathcal{T})-w^{*} < \frac{2(|V|-1)}{|E|\epsilon}\left(\left(|V|-1\right)\ln\left(\frac{|V|-1}{\gamma}\right) +\overset{|V|-1}{\underset{i=1}\sum} \ln|\mathcal{R}_{i}|\right) $$ with probability at least $1-\gamma$, and
with $w^{*} = \overset{|V|-1}{\underset{i=1}\sum} \min\limits_{r \in \mathcal{R}_{i}}w(r).$
\end{theorem}

\begin{preuve}
For clarity of the proof, let us denote $\epsilon' :=\frac{\epsilon}{|V|-1}$, $ OPT_{i} := OPT_{u_{\mathcal{G}}}\left(w_{|\mathcal{R}_{i}}\right)$,  $u_{i} :=$ $u_{\mathcal{G}}\left(w,\mathcal{M}_{Exp}\left(\mathcal{G},w,u_{\mathcal{G}},\mathcal{R}_{i},\epsilon'\right) \right)$, and finally $r_{i} \in \mathcal{R}_{i}$ the edge selected at the $i^{\text{th}}$ step of the algorithm.

According to Proposition~\ref{tradeoffexponential} one gets $\forall i \in [|V|-1]$
\begin{align*}
&\mathbb{P}\left[u_{i}\leq OPT_{i}-\frac{2\Delta u_{\mathcal{G}}}{\epsilon'}\left( t +\ln|\mathcal{R}_{i}| \right)\right] \leq e^{-t}. \\
\intertext{Then, by union bound, one gets}
&\mathbb{P}\left[\exists i \textit{ s.t } \left\{ u_{i}\leq OPT_{i}-\frac{2\Delta u_{\mathcal{G}}}{\epsilon'}\left( t+\ln|\mathcal{R}_{i}|\right)\right\} \right] \leq (|V|-1) e^{-t}.
\intertext{This probability is also straightforwardly an upper bound of} 
& \mathbb{P}\left[\overset{|V|-1}{\underset{i=1}\sum} u_{i}\leq  \overset{|V|-1}{\underset{i=1}\sum} OPT_{i} - \frac{2\Delta u_{\mathcal{G}}}{\epsilon'}\left( t+\ln|\mathcal{R}_{i}|\right)\right].
\intertext{Given the form of the chosen utility function, one gets $\forall i \in [|V|-1]$, $OPT_{i}=0$. Moreover $\Delta u_{\mathcal{G}} = 1/|E|$, thus } 
= & \mathbb{P}\left[\overset{|V|-1}{\underset{i=1}\sum} u_{i}\leq  - \frac{2}{|E| \epsilon'}\left(\left(|V| -1\right) t + \overset{|V|-1}{\underset{i=1}\sum} \ln|\mathcal{R}_{i}| \right) \right].
\intertext{Since  $u_{i} = \min\limits_{r \in \mathcal{R}_{i}} w(r) -w(r_{i})$ , on  gets $\overset{|V|-1}{\underset{i=1}\sum} u_{i} = w^{*} -\overset{|V|-1}{\underset{i=1}\sum} w(r_{i}) $, thus}
= & \mathbb{P}\left[w^{*} -\overset{|V|-1}{\underset{i=1}\sum} w(r_{i}) \leq  - \frac{2}{|E| \epsilon'}\left(\left(|V| -1\right) t + \overset{|V|-1}{\underset{i=1}\sum} \ln|\mathcal{R}_{i}| \right) \right]
\intertext{Setting $(|V|-1)e^{-t}=\gamma$, replacing $\epsilon'$, since $w(\mathcal{T}) =\overset{|V|-1}{\underset{i=1}\sum} w(r_{i})$, and using the equation (5) and (2) one gets}
& \mathbb{P}\left[\overset{|V|-1}{\underset{i=1}\sum} w(r_{i}) -w^{*} \geq  \frac{2(|V|-1)}{|E| \epsilon}\left(\left(|V| -1\right) \ln\left(\frac{|V|-1}{\gamma}\right) + \overset{|V|-1}{\underset{i=1}\sum} \ln|\mathcal{R}_{i}| \right) \right] \leq \gamma \\
\intertext{therefore, one finally gets}
& w(\mathcal{T})-w^{*} < \frac{2(|V|-1)}{|E|\epsilon}\left(\left(|V|-1\right)\ln\left(\frac{|V|-1}{\gamma}\right) +\overset{|V|-1}{\underset{i=1}\sum} \ln|\mathcal{R}_{i}|\right) \\
\intertext{with probability at least $1-\gamma$. \hfill\ensuremath{\blacksquare} } \notag\end{align*}\end{preuve}
This theorem gives the difference between the weight of the output tree topology, and the value $w^{*}$ which is the sum of the optimal choices at every step  of the algorithm. Since the steps are fixed, selecting the minimal weighted edge at every step does not necessarily produce a tree, this is why, to obtain a theoretical bound for the error on the weight of the minimum spanning tree of $G$, one need the following result.

\begin{prop}
\label{prop:w*lessMST}
Let $G=(V,E,w)$ the studied graph, if one denotes $w(\mathcal{T}^{*})$ the weight of the minimum spanning tree on $G$, given the above notation, one has $w^{*} \leq w(\mathcal{T^{*}}).$
\end{prop}

\begin{preuve}
Let $G=(v,E,w)$, $\mathcal{T}$ the minimum spanning tree of $G$ denoted by a set of edges $(e_{i})_{i\in \{1,...,|V|-1\}}$. One can always consider $w\left(e_{1}\right)< ... < w\left(e_{|V|-1}\right)$. We only consider the strict inequalities for readability of the proof, however a generalization of this proof to classical inequalities is straightforward due to the fact that treating the equality cases is immediate   . Let us also denote $(S_{i})_{i\in [|V|-1]}$ the sets of nodes at every step of the algorithm, $(\mathcal{R}_{i})_{i\in [|V|-1]}$ the corresponding xor-incident edges, and finally $\forall i \in [|V|-1], I_{i} = \underset{j \in [i]}{\cup}\mathcal{R}_{j}$ the sets of edges that are incident to one of the nodes in $S_{i}$. For readability of the proof, we also denote by $|.|_{\mathcal{T}}$ the number of edges from $\mathcal{T}$ that a set of edges contains.

\begin{lemma}
\label{lemma:graphproof}
For all $i$ in $[|V|-1]$, $\mathcal{R}_{i}$ contains at least one edge from $\mathcal{T}$, and $I_{i}$ contains at least $i$ edges from $\mathcal{T}$.
\end{lemma}

\begin{preuve}
Let us first suppose that $\exists i \in [|V|-1]$ such that $\mathcal{R}_{i}$ contains no edge from $\mathcal{T}$, then we managed to produce a cut of the graph $\left(S_{i},V\backslash S_{i}\right)$ such that no edge of $\mathcal{T}$ crosses it. By definition of a spanning tree, this is impossible. Therefore for all $i$ in $[|V|-1]$, $|\mathcal{R}_{i}|_{\mathcal{T}} \geq 1$.\\
Second, let us show by recurrence that for all $n \in [|V|-1]$
$P_{n}:=$ "$|I_{n}|_{\mathcal{T}} \geq n$" is true. \\
$P_{1}$ is true according to the first point of this theorem and since $\mathcal{R}_{1}=I_{1}$.
Given $k \in [|V|-2]$, let us suppose that $P_{k}$ is true.
If $I_{k+1}\backslash I_{k} =  \mathcal{R}_{k+1}\backslash I_{k} \neq \emptyset$ then $P_{k+1}$ is automatically satisfied.
Else let us suppose that $|I_{k}|_{\mathcal{T}} \leq k$, then $|I_{k+1}|_{\mathcal{T}} \leq k$. According to the first part of the proposition, $\exists q \geq 1$ such that $|\mathcal{R}_{k+1}|_{\mathcal{T}} = q$, therefore $|I_{k+1}\backslash \mathcal{R}_{k+1}|_{\mathcal{T}} \leq k-q $. Then the graph edge-induced by $I_{k+1}\backslash \mathcal{R}_{k+1}$ contains at least $q+1$ subtrees of $\mathcal{T}$. Since $|\mathcal{R}_{k+1}|_{\mathcal{T}} = q$, at least one subtree, denoted $S^{*}$, does not have an incident edge in $\mathcal{T} \cap \mathcal{R}_{k+1}$. We just produced a cut $\left(S^{*},V\backslash S^{*}\right)$ such that no edge of $\mathcal{T}$ crosses it. This contradiction implies that our assumption is false, then $|I_{k}|_{\mathcal{T}} = |I_{k+1}|_{\mathcal{T}} \geq k+1$. We just proved that $P_{k}$ implies $P_{k+1}$, therefore Lemma~\ref{lemma:graphproof} is proved by recurrence. \hfill\ensuremath{\blacksquare}
\end{preuve}

Let $k>1$, for all $j>k$, thanks to Lemma~\ref{lemma:graphproof} $|I_{j}|_{\mathcal{T}} \geq j$. At least $j-k+1$ edges in $I_{j}\cap\mathcal{T}$ have a weight lower or equal to  $w(e_{|V|-k})$ by definition of $(e_{i})_{i\in [|V|-1]}$, thus at least $j-k$ steps of the algorithm  had an optimal solution with weight lower or equal to $w(e_{|V|-k})$. In particular for $j=|V|-1$, at least $|V| - k $ steps of the algorithm had an optimal solution with weight lower or equal to $w(e_{|V|-k})$. Therefore, 
for all $k\geq 2$, their is at most $k-1$ sets $\{\mathcal{R}_{i_{1}},...,\mathcal{R}_{i_{k-1}} \}$ such that $\min\limits_{r\in \mathcal{R}_{i_{.}}} w(r) > w(e_{|V|-k})$.

Then, in the worst case scenario, $\forall k \in [2,|V|-1]$, there exists exactly $k-1$ steps such that $\min\limits_{r\in \mathcal{R}_{i_{.}}} w(r) > w(e_{|V|-k})$.
The only possible way to have such sets is if one has $\{i_{1},...,i_{|V|-1}\}$ such that $\min\limits_{r\in \mathcal{R}_{i_{1}}} w(r) > w(e_{|V|-2}) \leq \min\limits_{r\in \mathcal{R}_{i_{2}}} w(r) > w(e_{|V|-3}) ...  > w(e_{1}) \geq \min\limits_{r\in \mathcal{R}_{i_{|V|-1}}} w(r).$
Moreover, since $\forall i \in [V-1], |\mathcal{R}_{i}|_{\mathcal{T}} \geq 1$, then $\forall i \in [V-1], w(e_{|V|-1}) \geq \min\limits_{r\in \mathcal{R}_{i}} w(r).$ Finally, one has $w(e_{|V|-1}) \geq \min\limits_{r\in \mathcal{R}_{i_{1}}} w(r) > w(e_{|V|-2}) \leq \min\limits_{r\in \mathcal{R}_{i_{2}}} w(r) > w(e_{|V|-3}) ...  > w(e_{1}) \geq \min\limits_{r\in \mathcal{R}_{i_{|V|-1}}} w(r)$ therefore 
$w^{*}=\sum\limits_{i\in [|V|-1]}\min\limits_{r \in \mathcal{R}_{i}} w(r) \leq \sum\limits_{i\in [|V|-1]} w(e_{|V|-i}) = w(\mathcal{T}) \hfill\ensuremath{\blacksquare}$
\end{preuve}

To conclude on the theoretical worst case error of the algorithm, we introduce the following corollary, using the approximation error from Definition~\ref{definition:error}.

\begin{corollary}
\label{theoreticalbound}
Let $G=(\mathcal{G},w)$ be a weighted graph, $\epsilon >0$, and $\mathcal{T}= PAMST(\mathcal{G},w,\epsilon)$, then $$Error_{G}\left(\mathcal{T}\right) \leq \frac{2(|V|-1)}{|E|\epsilon}\left((|V|-1)\ln\frac{|V|-1}{\gamma} + 2
\ln \left( \left(V-1\right)!\right) \right) $$ with probability $1 - \gamma$.
\end{corollary}

\begin{preuve}
Since for any $\left \{\mathcal{R}_{1},..., \mathcal{R}_{|V|-1}\right\}$ the sets of xor-incident edges, one has $|\mathcal{R}_{i}|\leq i\left(|V|-i\right)$, thus $\overset{|V|-1}{\underset{i=1}\sum} \ln|\mathcal{R}_{i}| = \ln \left( \overset{|V|-1}{\underset{i=1}\prod} |\mathcal{R}_{i}|\right)\leq \ln \left(  \overset{|V|-1}{\underset{i=1}\prod} i\left(|V|-i\right) \right) = 2
\ln \left( \left(V-1\right)!\right)$. Therefore, this corollary is proved by binding the Theorem~\ref{firstbound} and Proposition~\ref{prop:w*lessMST}.\hfill\ensuremath{\blacksquare}
\end{preuve}

According to the above corollary, the theoretical bound is useful when the studied graph is dense. In fact, if $|E| \simeq V(V-1)/2$, the bound becomes $O\left(\ln(V-1)\right)$ for fixed privacy parameters.
Therefore, the presented algorithm is particularly suited to the study of dense networks, allowing to keep differential privacy for almost no approximation error. 
Given the achieved positive results, it is natural to compare this new In-processing method to the Post-processing Laplace mechanism. To do so, Section~\ref{section:boundcomp} produces a theoretical and empirical comparison of the error of both methods.

\section{Bounds comparison}

This section produces two theorems for comparing the two theoretical bounds (Theorem~\ref{theorem:Sealfonbound} versus Corollary~\ref{theoreticalbound}). The first one gives a bound according to the degree of certainty $\gamma$, and the number $|V|$ of nodes in the graph. The second one does not depend on the parameters.
First, let us introduce some notations, and recall the definition of the Lambert W-function.

\begin{deff}[Lambert W-function]
The Lambert W-function, denoted $W$, also called the omega function, is the inverse function of $f$ where
$$ \forall z \in \mathbb{C}, f(z)=ze^{z}$$
\end{deff}

Denoting $\mathcal{B}_{Exp,\epsilon,\gamma}$ the bound from Corollary~\ref{theoreticalbound}
and $ \mathcal{B}_{Lap,\epsilon,\gamma}$ the one from Theorem~\ref{theorem:Sealfonbound} the first comparison theorem is as follows.


\begin{theorem}
\label{theorem:tightbound}
Let $G=(V,E,w)$ be a weighted graph, 
given that we evaluate the bounds of the two mechanisms according to the same degrees of certainty and privacy $\gamma$ and $\epsilon$, 
we have $$\text{ if } \alpha \geq \frac{\gamma}{n}e^{W\left(\ln\left(h_{n,\gamma}\right)\right)} \text{ then } \mathcal{B}_{Exp,\gamma} \leq \mathcal{B}_{Lap,\gamma} $$
with $$h_{n,\gamma} := \gamma^{-\frac{n}{\gamma}}(2\pi)^{\frac{1}{\gamma}}n^{\frac{3n+1}{\gamma}}e^{-\frac{2n}{\gamma}+\frac{1}{6n\gamma}}$$

and with the following notations: $|E|$ being the number of edges in a connected graph and $|V|$ its number of nodes, one can always write $|E| = \alpha (|V|-1)$ with $\alpha \geq 1$ ($\alpha$ can be seen as a non-normalized density of the graph). Moreover for readability, we denote $(|V|-1)=n$.
\end{theorem}

\begin{preuve}
Let us first recall the following lemma:
\begin{lemma}\cite{Robbins1985}
\label{lemma:robbins}
$ \forall n \in \mathbb{N}^{*},$ $$ n! \leq \sqrt{2\pi} n^{n+\frac{1}{2}} e^{-n +\frac{1}{12n}}$$
\end{lemma} 

Let $G=(V,E,w)$ be a weighted graph.
Let us consider that we have fixed values for $\gamma$ and $\epsilon$.

One has that:
\begin{align*}
& \mathcal{B}_{Exp,\gamma} \leq \mathcal{B}_{Lap,\gamma}\\
\Leftrightarrow & \frac{2(|V|-1)}{|E|\epsilon}\left(\left(|V|-1\right)\ln\left(\frac{|V|-1}{\gamma}\right) + 2\ln\left(\left(|V|-1\right)!\right)\right) \leq \frac{2(|V|-1)}{\epsilon}\ln\left(\frac{|E|}{\gamma}\right) \\
\shortintertext{denoting $|E|=\alpha(|V|-1)$ and $(|V|-1)=n$ }  \notag \\
\Leftrightarrow & \frac{2n}{\alpha n\epsilon}\left(n\ln\left(\frac{n}{\gamma}\right) + 2\ln\left(n!\right) \right)\leq \frac{2n}{\epsilon}\ln\left(\frac{\alpha n}{\gamma}\right) \\ 
\Leftrightarrow & \frac{1}{\alpha n}\left(n\ln\left(\frac{n}{\gamma}\right) + 2\ln\left(n!\right)\right) \leq \ln\left(\frac{\alpha n}{\gamma}\right) \\ 
\shortintertext{Using Lemma~\ref{lemma:robbins}, it is sufficent to show: }  \notag \\
& \frac{1}{\alpha n}\left(n\ln\left(\frac{n}{\gamma}\right) + \ln\left(2\pi n^{2n+1} e^{-2n +\frac{1}{6n}}\right) \right) \leq \ln\left(\frac{\alpha n}{\gamma}\right) \\ 
\Leftrightarrow & \frac{1}{\alpha n}\ln\left(\left(h_{n,\gamma}\right)^{\gamma}\right) \leq \ln\left(\frac{\alpha n}{\gamma}\right) \\ 
\Leftrightarrow & \frac{\gamma}{\alpha n}\ln\left(h_{n,\gamma}\right) \leq \ln\left(\frac{\alpha n}{\gamma}\right) \\ 
\Leftrightarrow & \ln\left(h_{n,\gamma}\right) \leq \frac{\alpha n}{\gamma}\ln\left(\frac{\alpha n}{\gamma}\right)=g\left(\frac{\alpha n}{\gamma}\right)\numberthis \label{equation:tightbound}
\end{align*}

$g:= g(x) = x\ln(x)$ is a non-decreasing function on $\mathbb{R}^{+}$, thus solving  $\ln(h_{n,\gamma}) =g\left(\frac{\alpha n}{\gamma}\right)$ is sufficient to find the value $\alpha^{*}$ such that for all $\alpha$ bigger than this value, Equation~(\ref{equation:tightbound}) is satisfied.

To find such a solution it suffices to remark that $g(x)=\ln(x)e^{\ln(x)}$ thus using the Lambert W-function one gets $\ln\left(\frac{\alpha n}{\gamma}\right)=W\left(\ln(h_{n,\gamma})\right)$, therefore $\alpha^{*}=\frac{\gamma}{n}e^{W(h_{n,\gamma})}$.\hfill\ensuremath{\blacksquare}

\end{preuve}

If the second result is not as tight as the first one, it has the asset of depending neither on the order of the graph or any privacy/certainty degree.

\begin{theorem}
Let $G=(V,E,w)$ be a simple-undirected-weighted graph, 
Regardless of the degrees of certainty and privacy $\gamma$ and $\epsilon$, one has that $$ \alpha \geq 3 \implies \mathcal{B}_{Exp,\epsilon,\gamma} \leq \mathcal{B}_{Lap,\epsilon,\gamma} $$
\end{theorem}

\begin{preuve}
Let G=(V,E,w) be a weighted graph.
Let us consider that we have fixed values for $\gamma$,$\epsilon$.
One has that:
\begin{align*}
& \mathcal{B}_{Exp,\gamma} \leq \mathcal{B}_{Lap,\gamma}\\
\Leftrightarrow & \frac{2(|V|-1)}{|E|\epsilon}\left(\left(|V|-1\right)\ln\left(\frac{|V|-1}{\gamma}\right) + 2\ln\left(\left(|V|-1\right)!\right)\right) \leq \frac{2(|V|-1)}{\epsilon}\ln\left(\frac{|E|}{\gamma}\right) \\
\shortintertext{denoting $|E|=\alpha(|V|-1)$ and $(|V|-1)=n$ } 
\Leftrightarrow & \frac{2n}{\alpha n\epsilon}\left(n\ln\left(\frac{n}{\gamma}\right) + 2\ln\left(n!\right) \right)\leq \frac{2n}{\epsilon}\ln\left(\frac{\alpha n}{\gamma}\right)
\shortintertext{Since $n!\leq n^{n}$ it suffices to show } 
& \ln\left(\frac{n}{\gamma}\right) +2\ln(n)  \leq \alpha \ln\left(\frac{\alpha n}{\gamma}\right) \numberthis \label{equation:simplebound}
\shortintertext{
Finally, $\gamma \in (0,1]$,  $\ln\left(\frac{n}{\gamma}\right) \geq \ln(n)$, one thus easily gets that Eq.~(\ref{equation:simplebound}) is satisfied for any $\alpha \geq 3$. \hfill\ensuremath{\blacksquare}} \notag
\end{align*} 
\end{preuve}

\label{section:boundcomp}

\section{Experimental comparison of the methods}
\label{section:experiments}

\begin{table*}[!ht]
\centering
\begin{tabular}{|c||*{4}{c|}|c|}\hline
\backslashbox{p}{$\epsilon$}
&\makebox[3em]{0.1}&\makebox[3em]{0.4}&\makebox[3em]{0.7}
&\makebox[3em]{1.0}&\makebox{MST cost}\\\hline\hline
Laplace 0.1 & 4055.5$\pm$ 90.6&2191.2 $\pm$67&1301.9$\pm$42.9&876.4$\pm$30.5&\multirow{2}*{114-125}\\
\textsc{PAMST} 0.1 &322.3$\pm$12.5&45.7$\pm$3.1&16.8$\pm$1.4&8.5$\pm$0.8&\\\hline \hline
Laplace 0.3 &4139.3$\pm$77.3&2280.1$\pm$72.2&1384.5$\pm$41.1&965.0$\pm$31.8&\multirow{2}*{39-42}\\
\textsc{PAMST} 0.3 &108.7$\pm$ 3.8&15.2$\pm$1.0&5.6$\pm$0.5&2.8$\pm$0.3&\\\hline \hline
Laplace 0.5 &4152.8$\pm$84.0&2298.4$\pm$77.2&1396.2$\pm$47.8&975.7$\pm$29.7&\multirow{2}*{23-25}\\
\textsc{PAMST} 0.5 &64.7$\pm$2.5&9.1$\pm$0.6&3.4$\pm$0.2&1.7$\pm$0.2&\\\hline \hline
Laplace 0.7 &4151.0$\pm$95.2&2291.3$\pm$67.3&1400.4$\pm$42.5&979.6$\pm$31.0&\multirow{2}*{17-18}\\
\textsc{PAMST} 0.7 &64.7$\pm$2.5&9.1$\pm$0.6&2.4$\pm$0.2&1.2$\pm$0.1&\\\hline \hline
Laplace 0.9 &4159.6$\pm$82.6&2297.9$\pm$62.2&1408.3$\pm$44.2&983.8$\pm$32.8&\multirow{2}*{13-14}\\
\textsc{PAMST} 0.9 &36.2$\pm$1.6&5.0$\pm$0.3&1.9$\pm$0.2&0.9$\pm$0.1&\\ \hline 
\end{tabular} 
   \caption{Expected approximation error of Laplace and \textsc{PAMST} according to the Erd\"os Renyi probability for simulation (p), a degree of privacy ($\epsilon$) and confidence interval with probability 0.95. The size of the simulated graph is around $ \frac{p}{2}\times 10^{5}$, and an approximated interval of values (a-b) of the 
minimum spanning tree real cost for every density is indicated (MST cost).}
   \label{experiment:recapitulatiftable}
 \end{table*}

\begin{figure}[!ht]
\centering
\includegraphics[width=0.9\textwidth]{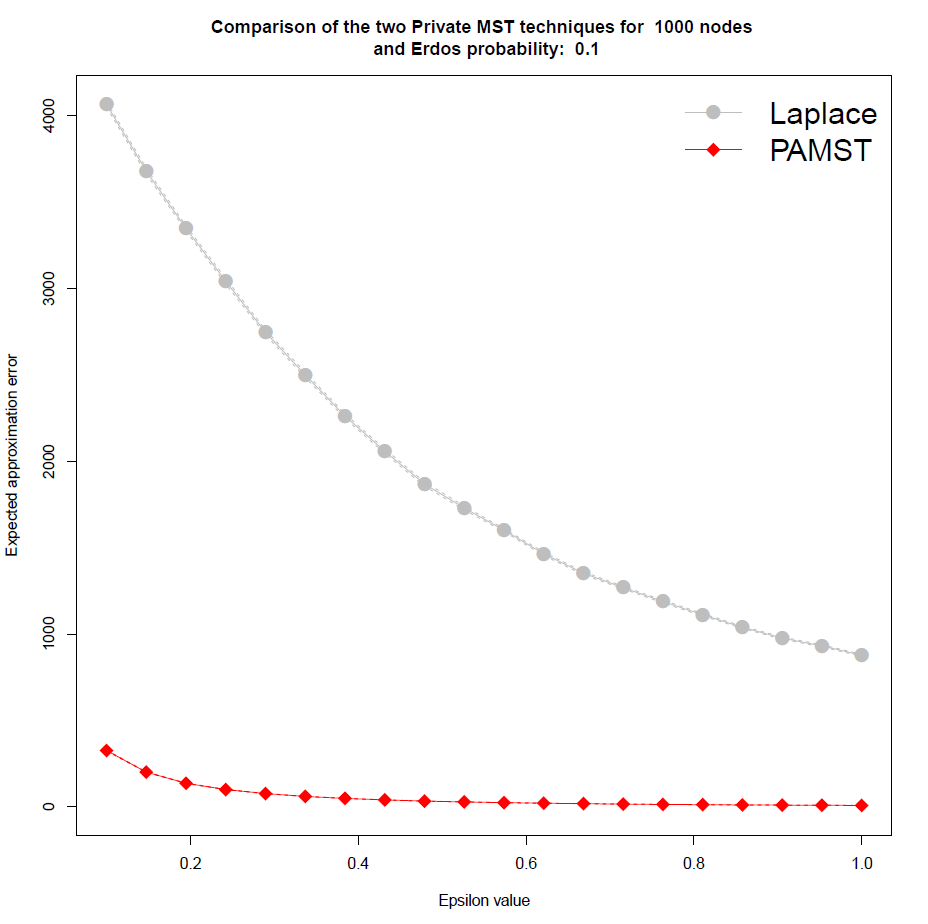}
\caption{Evaluation of the expected approximation error of the tree produced by post -processing Laplace mechanism and \textsc{PAMST} with confidence intervals with probability 0.95 (unreadable because too close to the curves) }
\label{figure:experimentdensity}
\end{figure}

To conclude this analysis of \textsc{PAMST} algorithm, we conducted some experiments on simulated graphs to visualize the improvement of \textsc{PAMST} in comparison to a post-processing of the Laplace mechanism. 
The following experiments investigate the expected approximation error for the spanning tree topologies produced rather by \textsc{PAMST} or the Laplace mechanism. Formally, it estimates $\mathbb{E}\left(Error_{G}\left(\mathcal{T}_{PAMST}\right) \right)$, and $\mathbb{E}\left(Error_{G}\left(\mathcal{T}_{Laplace}\right) \right)$ with the obvious notation $\mathcal{T}_{M}$ the tree produced by the algorithm $M$ and with the probability space being on the simplex of the mechanism. Figure~\ref{figure:experimentdensity} represents the expected approximation error of the two mechanisms according to a degree of privacy $\epsilon \in (0,1]$ for a graph Erd\"os-Renyi graph with probability 0.1.
The expected error calculus is based on computing the mean value of the spanning tree error for each tree method for the simulation of a hundred graphs. For every graph, the weight of each edge is randomly sampled from a standard uniform distribution $\mathcal{U}(0,10)$.The expected approximation error is thus produced according to a Monte-Carlos type estimation, with for each point a confidence interval with probability 0.95. As expected from the theoretical bounds, \textsc{PAMST} widely outperforms the Post-processing Laplace mechanism. Figure~\ref{figure:experimentdensity} is only a visualization of the shape of the two errors functions, Table~\ref{experiment:recapitulatiftable} represents a more extensive study of both of those expected errors for several densities and privacy parameters.
The analysis of this table highlights some interesting points about both methods.
First, the Post-processing of the Laplace mechanism do not adapt to the density of the graph, which is mostly due to the fact that the Laplace mechanism adds random noise to the edges, regardless of the number of edges in the graph since the sensitivity is set to one.
Second, the weight are simulated from a uniform distribution, therefore, the number of edges with small weights increase uniformly with the density of the graph. This implies that at every iteration of the exponential mechanism, the probability to sample an edge with close-to-optimal weight increase.

Such a great accuracy does not come without any cost. Our new method is a strong tool to ensure privacy and accuracy, however, sampling from a new distribution at every stage of the algorithm is quite expensive in terms of computation time. Producing a tree according to the \textsc{PAMST} method on a graph of size $5\times10^{5}$ takes about 40 seconds (Intel\textregistered  Xeon\textregistered  E5-2667 V4 3.20 GHz processor), which is quite faster than a Kruskal algorithm (efficient C++/Boost implementation takes more than a minute) but way slower than a prim algorithm (efficient C++ implementation takes less than 1 second). This contrast is mostly due to the use of priority queue in the implementation prim algorithm. This kind of implementation is not allowed for \textsc{PAMST} since the exponential mechanism should be independently computed at each step of the algorithm, and since there is no trivial way of computing an accurate priority queue for the whole graph. \\

\section{Discussion}
\label{section:discussion}

First, \textsc{PAMST} algorithm is presented here in a global framework where the maximal sensitivity of every edge weight is $1/2$. This choice, as well as the choice of normalization is made so that our method can compare to the former one ~\cite{Sealfon_2016}. We produced results according to this framework, however every results are scalable to any real factor. This means that if we multiply the weights sensitivity or the degree of privacy by a factor $M$, the bound presented in Corollary~\ref{theoreticalbound} will also scale to $M$.
Second, the main conception difference between \textsc{PAMST} and the former methods is that the privacy mechanism is directly in the spanning tree construction (In-processing) while until now, the release of such a statistic on the graph has only been allowed by Post-processing of a more general sanitizing, for example, the Laplace mechanism.
Sanitizing all the weights has the advantage of allowing the analyst to release any information on the Post-processing according to the Proposition~\ref{postprocessing}, this is why, while producing a minimum spanning tree topology according to the Laplace mechanism, the analyst can automatically release the noisy-weights of the selected edges. However, sanitizing all the graph weights with the Laplace mechanism is costly in terms of accuracy.
\textsc{PAMST} is an algorithm that releases only the topology of the spanning tree. It selects every edge according to the exact weight function of the graph. The choice of releasing, at first, only the topology of the spanning tree is a winning strategy in terms of accuracy, but since the weight could be useful for any post-processing of the spanning tree, one may want to use a classical numeric mechanism to release a noisy version of those weights. Since we only sanitize the weights of the tree instead of the whole graph weights, the amount of noise needed is much lower especially for dense graphs.
Finally, releasing the tree weights with $\epsilon$-differential privacy makes the global release of the weighted spanning tree $2\epsilon$-differentially private, however, given the composition theorem and the scalability of our method one can choose to rescale the parameters of the two mechanisms to make the  weighted tree release $\epsilon$-differentially private with only a small accuracy loss.

Chapter~\ref{section:clustering} introduces a new way of conceiving differentially private clustering which is differentially private MST-based clustering. Two way of constructing such an analysis are presented and tested on several toy datasets.

\chapter{Application to private node clustering}
\label{section:clustering}

Graph clustering~\cite{Schaeffer07} appears to be a key tool for understanding the underlying structure of many data sets by locating nodes groups ruled by a specific similarity. 
The minimum spanning tree is known to help recognizing clusters with arbitrary shapes in MST-based clustering algorithms and thus can be used for wider applications than community detection.
This kind of structure adapted clustering could pave the way to mutual interaction analysis in genomics, proteomics, or even web analysis where privacy-preserving is not an option but a strong requirement. Chapter~\ref{section:clustering} aims at providing a new field in private data analysis denoted node clustering in a graph under weight differential privacy. Although clustering using differentially private has already been studied (\cite{Nissim_2007,Zhanglong_Ji_2014} ), to our knowledge, this is the first time node clustering under weight-differential privacy is investigated. Section~\ref{section:Motivation} provides some motivations for the use of MST-based clustering, while the remaining ones will present how to produce an MST-based differentially private node clustering algorithm and some preliminary experimental results.

\section{Motivation for MST-clustering and DBMSTClu}
\label{section:Motivation}

The minimum spanning tree is surely the most famous sparse representation of a weighted graph. MST-based algorithms rely on the idea that the structure of a graph is well represented by its MST. Xu et al. produced Claim~\ref{claim:motivation} tried to formalize this idea.

\begin{claim}[\cite{Xu_2002}]
\label{claim:motivation}
If one takes two points $c_{1}$, $c_{2}$ of a cluster $C$, then all data points in the tree path connecting $c_{1}$ and $c_{2}$ in the MST must be in $C$.
\end{claim}

To justify the use of any MST-based, we first introduce a specific definition of the separability condition of a cluster enriching~\cite{Xu_2002}, and second give a proof for Claim~\ref{claim:motivation}

\begin{deff}[Separability condition of a cluster]
Let $G=(V,E,w)$ a simple-undirected-weighted graph, $(V,d)$ be a metric space defined in $G$ according to the minimal-weighted path between nodes, and $D\subset V$ a dataset. $C\subset D$ is called a cluster if and only if for any partition $C=C_{1}\cupdot C_{2}$ one has
$$\arg \min\limits_{v \in D-C_{1}} \left\{ \min\left\{d(v,c) | c \in C_{1} \right\} \right\} \in C_{2}.$$
\end{deff}

\begin{preuve}[of Claim~\ref{claim:motivation}]
Let $c_{1}$, $c_{2}$ two points of a cluster $C$. Let denote $\mathcal{T}_{D}$ the MST of the graph induced by $D$ in $G$ (one can suppose it is connected for clarity).
Since the minimum spanning tree is a tree, their is an only path between $c_{1}$ and $c_{2}$ in $\mathcal{T}_{D}$.
Moreover since the graph is simple one can characterize this path by the succession of nodes it goes through. Let suppose there exists $v \in D$ in such a path such that $v$ is not in $C$.
Let $C_{1}$ be the maximal connected component of $\mathcal{T}_{D}$, containing $c_{1}$, for which all nodes are in $C$, and $C_{2}=C-C_{1}$. $C_{2}$ is not empty since at least $c_{2} \in C_{2}$, therefore $C=C_{1}\cupdot C_{2}$.
By construction of $C_{1}$, its nearest neighbor cannot be in $C$ and thus it is not in $C_{2}$. 
This contradiction gives us that $C$ is not a cluster and therefore the claim is proved by contraposition. \hfill\ensuremath{\blacksquare}
\end{preuve}

\begin{figure}[!ht]
\centering
\includegraphics[width=0.8\textwidth]{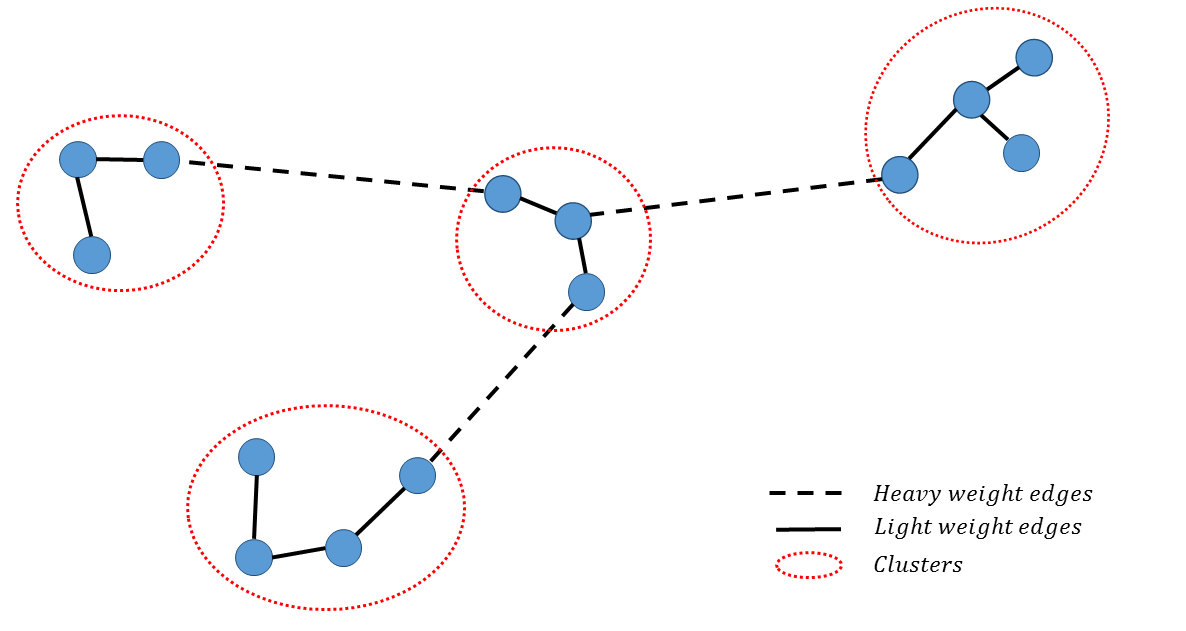}
\caption{Visualization of the general method for MST-based node clustering }
\label{figure:MSTclustering}
\end{figure}

This claim seams to be a good theoretical motivation for the use MST-based clustering. After investigating the most famous MST based algorithm ($MSDR$~\cite{Zhou_2011}) to have a global understanding of this kind a algorithms, we focused on a new method for MST-based node clustering denoted \textsc{DBMSTClu} introduced by Morvan and al. in~\cite{2017arXiv170302375M} and presented in Algorithm~\ref{algo:DBMSTClu}. The choice of this algorithm is mostly due to it simplicity (it does not need any parameter adjustments), its efficiency (refer to~\cite{2017arXiv170302375M}) and the ability to have an iterative and useful exchange with the authors.
As illustrated by Figure~\ref{figure:MSTclustering}, and like most of the MST-based clustering algorithm \textsc{DBMSTClu} starts by computing the minimum spanning tree of the graph, before performing some cuts among the edges of the tree according to a criteria on the edge weights. It aims, after several steps of well chosen cuts, at releasing a forest where each tree represents a cluster (the nodes that are in the same tree form a cluster).
To give a good understanding of the splitting criteria for \textsc{DBMSTClu} one must first introduce the following definitions.

\begin{deff}[Cluster Separation]
The Separation of a cluster $C_{i}$, denoted $SEP(C_{i})$, is defined as the minimum distance between the nodes of $C_{i}$, and the ones of all other clusters $C_{j}$, $i\neq j$,$1\leq i,j\leq K$, where $K$ is the total number of clusters. In practice, it corresponds to the minimum weight among all already cut edges from $\mathcal{T}$ incident to a node from $C_{i}$. If $K=1$, the Separation is set to 1. More formally, with $S_{j}$ the subtree of $\mathcal{T}$ corresponding to $C_{j}$,

\begin{equation*}
  \forall i \in [K],  \textit{ }SEP(C_{i}) = \left\{
      \begin{aligned}
\min\limits_{j \in [K], c_{j} \in Cuts(C_{i}) }  &w(c_{j}) &\textit{if } K\neq 1\\
&1 &otherwise.
      \end{aligned}
    \right.
\end{equation*}
\end{deff}

High values for Separation means that the cluster is well separated from the other clusters while low
values suggest that the cluster is well connected to other clusters.

\begin{deff}[Cluster Dispersion]
The Dispersion of a cluster $C_{i}$, denoted $DISP(C_{i})$, represented by the subtree $S_{i}$ of $\mathcal{T}$ is defined as the maximum edge weight of $S_{i}$. If the cluster is a singleton (i.e only contain one node), the corresponding Dispersion is set to 1.  More formally, with $S_{j}$ the subtree of $\mathcal{T}$ corresponding to $C_{j}$,
\begin{equation*}
  \forall i \in [K],  \textit{ }DISP(C_{i}) = \left\{
      \begin{aligned}
\min\limits_{ e_{j} \in S_{j} }  &w(e_{j}) &\textit{if $C_{j}$ is not a singleton}\\
&0 &otherwise.
      \end{aligned}
    \right.
\end{equation*}
\end{deff}

The validity index of a cluster and the validity index of a clustering partition are based on a normalized comparison of those to criterion and are the keystone to criterion for the $DBMSTCLu$ algorithm

\begin{deff}[Validity index of a cluster]
Let $i \in [K]$, and $C_{i}$ a cluster of nodes, the validity index of $
C_{i}$ is defined as $$V_{C}(C_{i}) = \frac{
SEP(C_{i})-DISP(C_{i})}{\max(SEP(C_{i}),DISP(C_{i}))} $$
\end{deff}

\begin{deff}[Validity index of a clustering partition]
The validity index of a partition $\Pi= \{C_{i}\}_{i \in [K]}$, denoted $DBCVI(\Pi)$, is defined as the weighted average of the validity indices of all the clusters in the partition. denoting $N$ the number of of points in the dataset, one has $$DBCVI(\Pi)= \sum \limits_{i \in [K]} \frac{|C_{i}|}{N}V_{C}(C_{i}) $$
\end{deff}

\begin{algorithm}[!ht]
\caption{\textsc{DBMSTClu}$(\mathcal{T},w)$~\cite{2017arXiv170302375M}}
\label{algo:DBMSTClu}
\begin{algorithmic} 
\REQUIRE A tree-topology $\mathcal{T}$, a weight function $w$. 
\ENSURE A list of node clusters represented by a list of subtrees of $\mathcal{T}$.
\STATE
\STATE $splitDBCVI \leftarrow -1$
\STATE $cut\_candidate\_list \leftarrow [ edges(T) ]$
\STATE $clusters = []$
\WHILE{$splitDBCVI<1$ }
\STATE $optimalcut \leftarrow None$
\STATE $tempDBCVI \leftarrow splitDBCVI $
\FOR {each $cut$ in $cut\_candidate\_list$}
\STATE $newClusters \leftarrow performCut(clusters,cut)$ 
\STATE $newDBCVI \leftarrow getDBCVI(newclusters,\mathcal{T})$
\IF {$newDBCVI \geq tempDBCVI $}
\STATE $optimal\_cut \leftarrow cut$
\STATE $tempDBCVI \leftarrow newDBCVI$
\ENDIF
\ENDFOR
\IF{ $optimal\_cut \neq None$}
\STATE $clusters \leftarrow performCut(clusters,optimal\_cut)$
\STATE $splitDBCVI \leftarrow getDBCVI(clusters,\mathcal{T})$
\STATE remove $optimal\_cut$ from $cut\_candidate\_list$
\ELSE 
\STATE $splitDBCVI \leftarrow 1$
\ENDIF
\ENDWHILE

\RETURN $clusters$
\STATE   
\end{algorithmic}
\end{algorithm}
\textbf{Additionnal notations:} performCut is the routine performing the cut corresponding to the edge in parameter with respect to already formed clusters. getDBCVI is the function computing the Validation Index of
the Clustering partition in parameter according to the initial MST.\\

\textsc{DBMSTClu} is summarized in Algorithm~\ref{algo:DBMSTClu}. It starts from a partition
with one cluster containing the whole dataset, and initializes the current value of DBCVI to -1 (the worst possible value). While there exists a cut which makes the current DBCVI greater, \textsc{DBMSTClu} greedily chooses the cut which maximizes the obtained DBCVI among all the possible cuts. When no direct improvement is possible, the algorithm stops. It returns the partition for which no improvement has been found. The main strength of this algorithm is that it is not parametric, therefore trying to construct a private algorithm based on this method will not force us to do a complex and tiresome parameter search. How to transform this deterministic algorithm for node clustering in a graph into a weight-differentially private 
method is thus what one should aim. As explained in Section~\ref{section:InprocessingVSPostprocessing} their is two main ways of constructing a private method named In-processing and Post-processing. For MST-based clustering the Post-processing method could rely on the construction of a private minimum spanning tree as a Pre-processing of a deterministic MST-based algorithm such as \textsc{DBMSTClu}. However In-processing methods are known to be strongly accurate, we chose to investigate this Post-processing method, given the work we produced on weight differentially private MST construction, and given that such a method will fit to any MST-based algorithm, leaving the In-processing \textsc{DBMSTClu} clustering for future work.

Section~\ref{section:Privateclustering} provides some simple theoretical conditions on the separability of the node clusters such that clustering based on a weight-differentially private approximated minimum spanning tree will not be too inaccurate, and experiments comparing the results of Private \textsc{DBMSTClu} based on the Laplace private MST or based on a \textsc{PAMST} output.

\section{Post-processing private MST-based clustering}
\label{section:Privateclustering} 

To introduce privacy in any MST-based node clustering, one can choose to produce first a approximated minimum spanning tree under weight differential privacy and then use this spanning tree in any MST-based algorithm. This Post-processing of a private approximated minimum spanning tree has the asset of being adaptable to any kind of clustering using the MST. This work only presents experiments for a private \textsc{DBMSTClu}, although this method can be trivially generalized to any MST-based algorithm. 

Chapter~\ref{section:PrivateMSTs} presented and compared two ways of producing an approximated minimum spanning tree under weight-differential privacy constraints. \textsc{PAMST} has been proven to be a much more accurate method in most cases, however the following theorem gives some conditions for the graph-based Laplace mechanism to sufficiently preserve the structure of the graph such that a MST-based algorithm using the Laplace Post-processing MST should produce an efficient clustering.

\begin{theorem}
\label{theorem:laplacecondition}
Let $G=(V,E,w)$ be an weighted graph,
if one constructs $K > 1$ sets of edges $(E_{1},...,E_{K})$ such that: \begin{align*} i,j \in [K], i<j \implies \forall e \in E_{i}, e' \in E_{j}, w(e) < w(e') \numberthis \label{equation:clusteringcondition} \end{align*} 
Then Eq.~(\ref{equation:clusteringcondition}) holds on $G'=\mathcal{M}_{GbL}(G,\epsilon)$  with probability greater than $$1-\sum\limits_{i \in [K-1]} \exp(- t_{i,i+1}\epsilon)\left(\frac{1}{2}+\frac{ t_{i,i+1} \epsilon}{4}\right)(E_{i}+E_{i+1}-1) \textit{ With } t_{i,i+1}=\min\limits_{e \in E_{i}, e' \in E_{i+1} } \left\{w(e')-w(e)\right\}.$$ 
\end{theorem}

\begin{preuve}
Let $G=(V,E,w)$ be a weighted graph, and $(e_{i})_{i \in \left[|E|\right]}$ the set of weight-ordered edge of $G$ such that $\forall i \in \left[|E|\right], w(e_{i}) \leq w(e_{i+1})$.
Let $G'=G=(V,E,w')=\mathcal{M}_{GbL}(G,\epsilon)$, one has $\forall i \in \left[|E|\right], w'(e_{i})=w(e_{i}) + Y_{i}$ with $Y_{i} \sim Lap(1/\epsilon).$ 
Let us give what the probability that the order has been preserved is, i.e 
\begin{align*}
&\mathbb{P}\left( w'(e_{1}) \leq w'(e_{2}) \leq ... \leq w'(e_{n})\right)
\shortintertext{
To begin with, one can consider $Y_{1}, Y_{2} \overset{iid}\sim Lap(1/\epsilon)$. Thus one has by symmetry of the Laplace distribution that }
&\mathbb{P}\left( w(e_{1}) + Y_{1} \leq w(e_{2}) + Y_{2} \right) = F_{X_{1}+X_{2}}\left(w(e_{2})-w(e_{1})\right)
\shortintertext{Moreover, one can easily show that}
&Y_{1},Y_{2}  \overset{iid}\sim Lap(1/\epsilon) \implies f_{Y_{1}+Y_{2}}= \frac{\epsilon}{4}\left(1+\epsilon |u|\right) \exp\left(- \epsilon |u|\right)
\shortintertext{Therefore one gets}
&\forall t >0 ,F_{Y_{1}+Y_{2}}(t) = 1-\exp(\epsilon t)\left(\frac{1}{2}+\frac{t \epsilon}{4}\right) \numberthis \label{equation:pdfofsumlaplace}
\shortintertext{Using the union bound and the iid property one gets}
&\mathbb{P}\left(\exists i,i+1 \text{, s.t }  w'(e_{i}) > w'(e_{i+1})\right)
\leq \left(1-F_{Y_{1}+Y_{2}}\left(\min\limits_{i \in [|E|-1]}\left\{w(e_{i})-w(e_{i+1})\right\}\right)\right)\times \left(|E|-1\right) \numberthis \label{equation:proofLaplacepreservestructure}
\shortintertext{In particular, if one wants to preserve the block structure between $K>1$ sets $(E_{1},...,E_{K})$, one can easily extend Eq.~(\ref{equation:proofLaplacepreservestructure}) to Eq.~(\ref{equation:clusteringcondition}) using Eq.~(\ref{equation:pdfofsumlaplace}) \hfill\ensuremath{\blacksquare} } \notag
\end{align*}
\end{preuve}

\begin{figure}[!ht]
\centering
\includegraphics[width=0.9\textwidth]{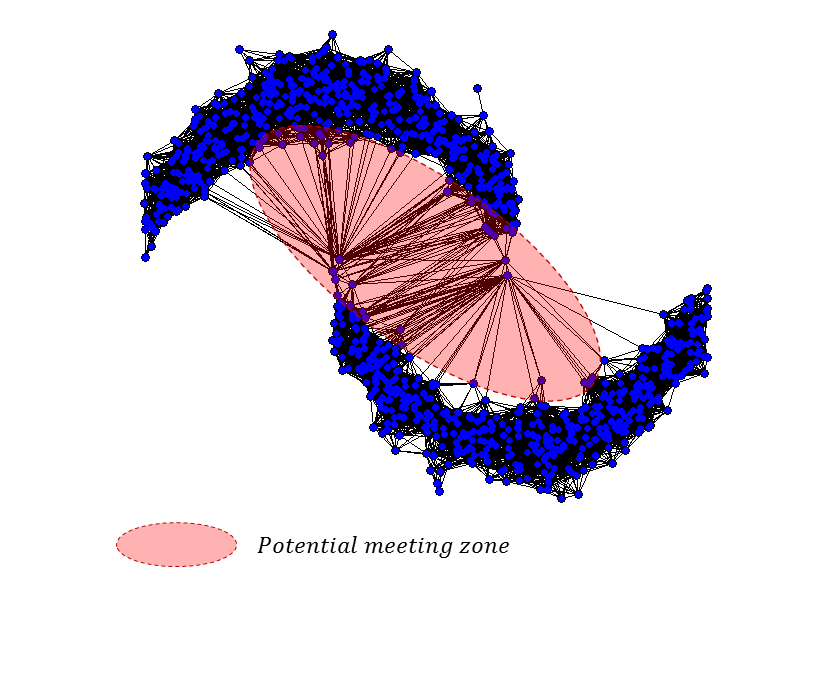}
\caption{Visualization of two node clusters, based on the euclidean distance between the nodes. The potential meeting zones represents the edges sensitive to the addition of noise. Those are the edges that may cause the clusters to collapse.}
\label{figure:Laplaceclusteringcondition}
\end{figure}

\begin{remark}
One will found in Appendix~\ref{appendixcomplementaryprooflaplacecondition} of this document the demonstration of the density of a sum of independent Laplace laws, as well as their distribution function for any $ t>0$.
\end{remark}

As illustrated by Figure~\ref{figure:Laplaceclusteringcondition} this theorem is much more of a local condition. In fact, if a dataset only has a few edges that are sensitive to noise injection (represented in Figure~\ref{figure:Laplaceclusteringcondition} by the nodes connected by an edge in the Potential meeting zone), Theorem~\ref{theorem:laplacecondition} will provide some theoretical guarantees on the clustering. For example if one wants to preserve $1000$ well chosen weighted edges in an spanning tree construction with 1-differential privacy, given that the minimum weight of an edge in the meeting zone is 10, those weights will be preserved with probability higher than $1-\exp(- 10)\left(\frac{1}{2}+\frac{10}{4}\right)(1000-1) \simeq 0.86$. However as soon as the number of distances one wants to preserve becomes too numerous, the privacy becomes too strong, or the distance between clusters too small, Theorem~\ref{theorem:laplacecondition} is quickly overstepped. This is why one could think of a private MST-based clustering using an approximated minimum spanning tree from the Post-processing Laplace method only when the clusters are well separated, and the privacy parameter close to 1. Relying on Section~\ref{section:boundcomp}, we claim that clustering based on an output of \textsc{PAMST} should be at least as accurate as the one from a Post-processing Laplace method.

Section~\ref{section:clusteringexperiments} presents an empirical investigation of this claim showing that private clustering using an approximated minimum spanning tree produced by \textsc{PAMST} is a more robust way to obtain clustering algorithm under differential privacy constraints.
\newpage

\section{Clustering experiments}
\label{section:clusteringexperiments}

\begin{figure}[t]
\centering
\includegraphics[width=0.6\textwidth]{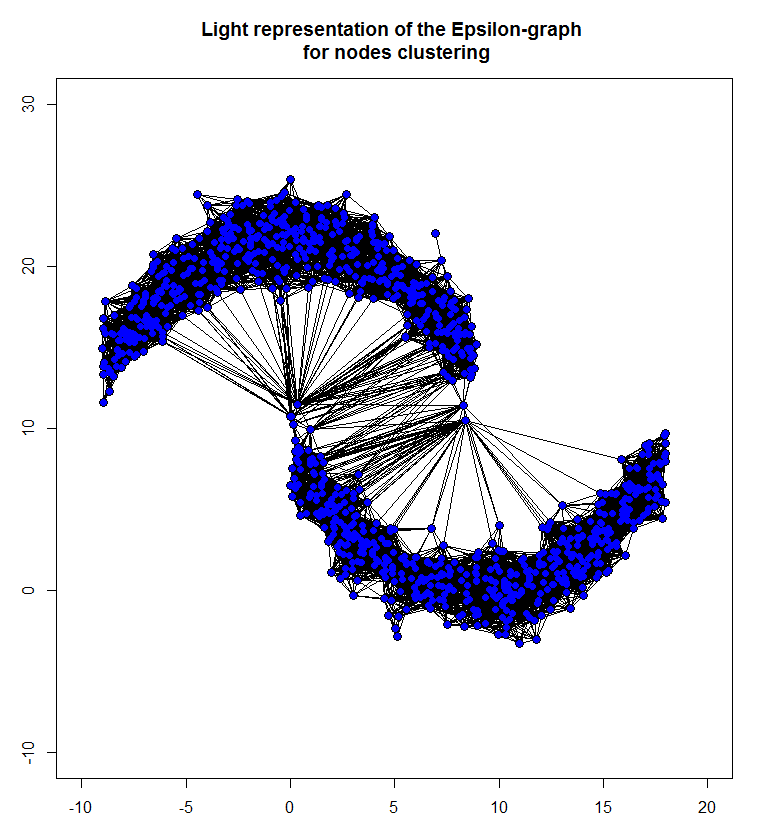}
\caption{Simple visualization of the $\epsilon$-graph based on the euclidean matrix distance of the simulated dataset}
\label{figure:epsilongraph}
\end{figure}

We conducted some experiments on arbitrarily shaped datasets to compare the two methods evoked in Section~\ref{section:Motivation} for nodes clustering under differential privacy constraints.
For this purpose, some datasets has been simulated based on $1000$ uniform point sampling over arbitrary shapes, perturbed by some random Gaussian noise. Given that the privacy definitions scale to $1$, we chose to also scale the variation of this random noise to $1$.
Each simulated datasets is composed of well separated clusters, and therefore, constitute a suitable benchmark for the evaluation of any clustering method. Although those datasets are not initially graphs, as evoked in Figure~\ref{figure:Laplaceclusteringcondition}, the euclidean distance matrix between all node pairs can be interpreted as complete weighted graphs and thus MST-based clustering can be performed. Moreover, the gap between the accuracy \textsc{PAMST} and the Laplace Post-processing increases accordingly with the density of the graph, we choose to narrow it down by computing an $\epsilon$-graph of the distance matrix. Figure~\ref{figure:epsilongraph} gives a visualization of this graph for one of the dataset we simulated.
For the following, we omit to represent the edges of this dataset for readability of the figures.
Finally, according to Section~\ref{section:discussion} and since \textsc{DBMSTClu} needs the to use the weights of the tree to produce a partition of the dataset, one has to decompose the releasing of a weighted approximated minimum spanning tree into two parts. Therefore, we firstly produce the topology of the spanning tree using \textsc{PAMST} under 1/2-differential privacy, and secondly add random noise to ensure 1/2-differential privacy on the weights of this tree-topology. 
\vspace{1cm}

\begin{figure}[t]
\centering
\includegraphics[width=0.45\textwidth]{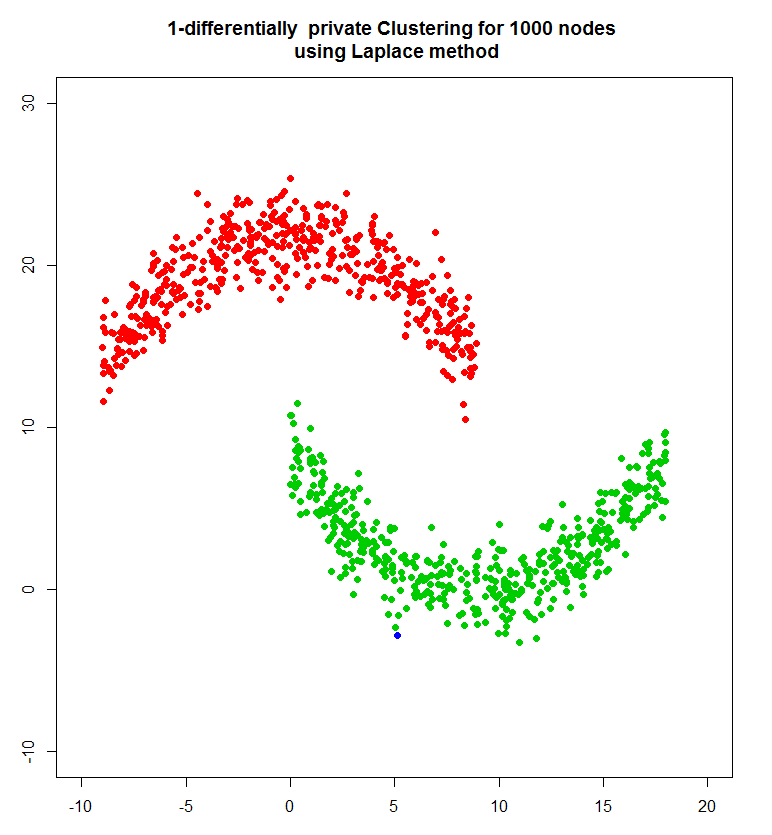}
\includegraphics[width=0.45\textwidth]{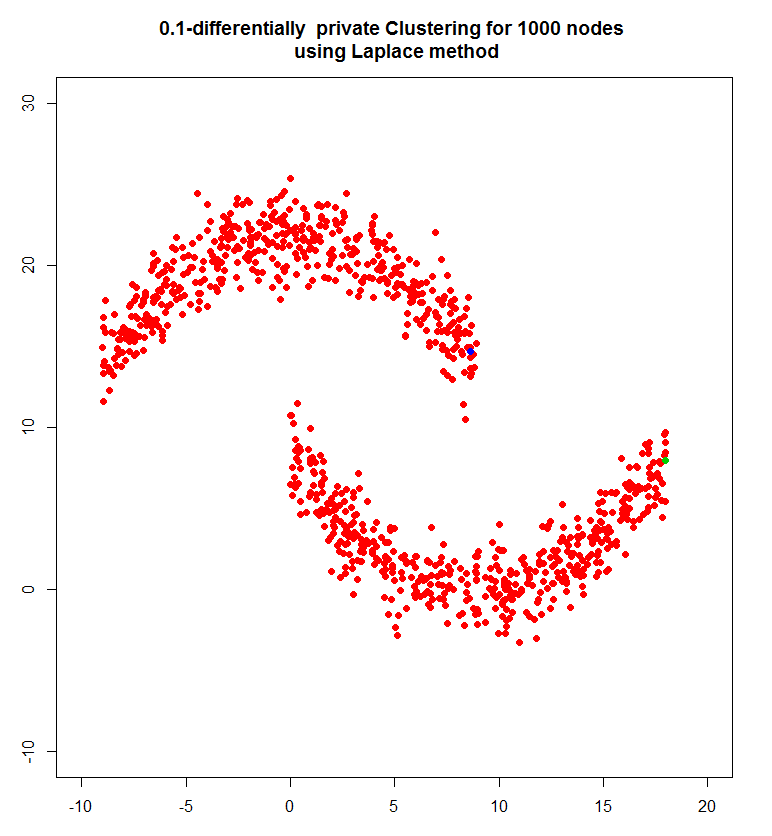}
\caption{Clustering under differential privacy on simulated dataset using Laplace sanitizing}
\label{figure:Lapsanitizclustering}
\end{figure}

\begin{figure}[t]
\centering
\includegraphics[width=0.45\textwidth]{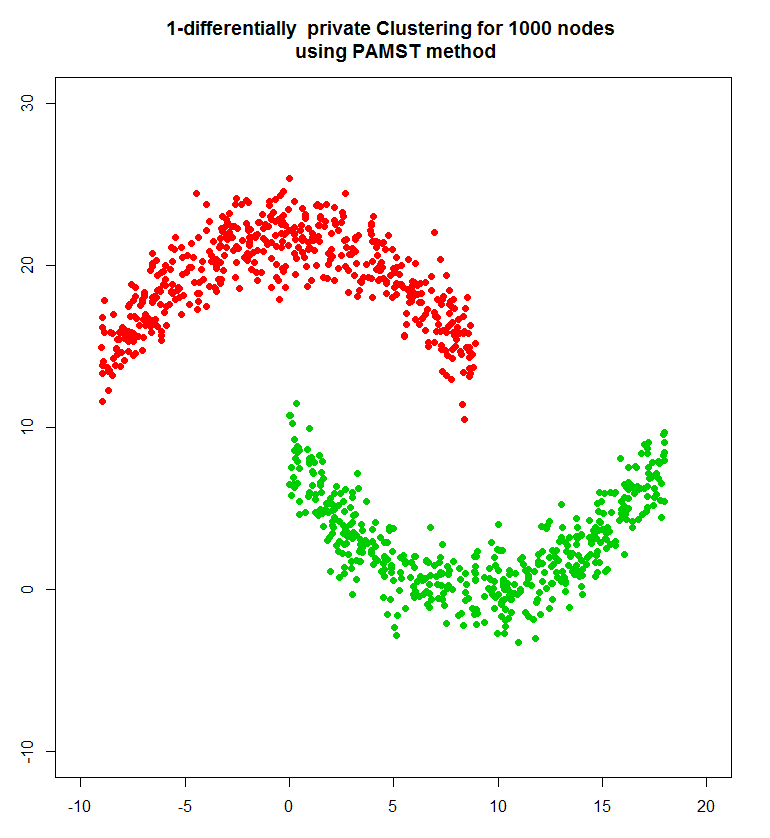}
\includegraphics[width=0.45\textwidth]{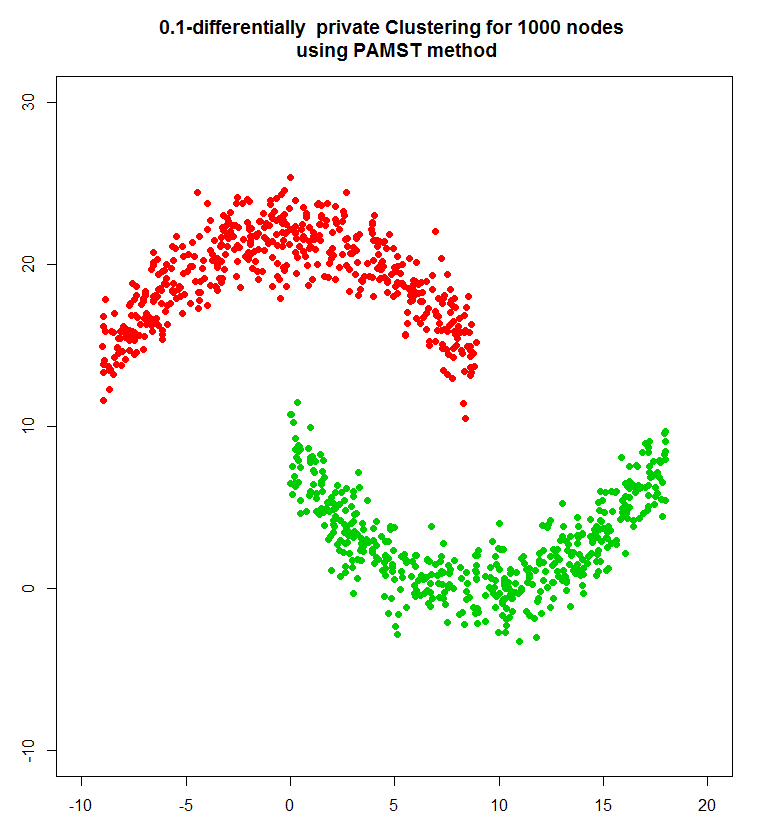}
\caption{Clustering under differential privacy on simulated dataset using \textsc{PAMST}}
\label{figure:PAMSTsanitizclustering}
\end{figure}

As expected according to Chapter~\ref{section:clustering}, when trying to ensure 0.1-differential privacy, The Laplace Post-processing method is not sufficiently precise to prevent the clustering algorithm from failing. On the other hand \textsc{PAMST} by producing a more accurate approximated minimum spanning tree allows \textsc{DBMSTClu} to present a convincing partition of the dataset. 
\newpage

\section{Discussion}
\label{section:dicussion2}
This chapter presented a new way of conceiving differentially private clustering. In fact, to our knowledge, this is the first time one can produce a MST-based clustering while ensuring differential privacy in the process. Moreover, we theoretically motivated MST-clustering as a good way of understanding the underlying structure of a dataset, while giving a clear proof for Claim~\ref{claim:motivation} initially stated in~\cite{Xu_2002}.
We also produced an experimental analysis to support the idea that private node clustering using \textsc{PAMST} outperforms the Laplace-based method.
As already stated in Section~\ref{section:experiments}, one should not overlook the computational cost of \textsc{PAMST} (one can refer to Section~\ref{section:experiments} keeping in mind that the Graph based Laplace mechanism is almost free in terms of computational cost).
Unfortunately, \textsc{DBMSTClu} is also quite costly since it is a greedy algorithm. Hopefully, thanks to parallel computing, this difficulty can be bypassed. We thus implemented \textsc{DBMSTClu} algorithm in \textsc{R} high-level programming language, using the library \textit{foreach} for a simple and efficient parallelization, and calling our \textsc{C++} implementation of \textsc{PAMST} using \textsc{Rcpp} compilation.

Finally, the analysis we gave in the last section is of course not supported only by this experiment. One can find some additional experiments on several others simulated datasets in Appendix~\ref{appendix:simu}.
One should however remark that those experiments, as well as the one in Figures~\ref{figure:Lapsanitizclustering}, and~\ref{figure:PAMSTsanitizclustering} are only fixed views of successful/unsuccessful clustering tasks. A different randomization could always cause our conclusion to be gainsaid, this is why for further work, we will focus on giving theoretical guarantees on the accuracy of such private clustering algorithms, as well as producing more exhaustive experiments one simulated and real-life datasets. 

\openany
\chapter*{Conclusion}
\addcontentsline{toc}{chapter}{Conclusion, publications, future work}
To conclude, in this report, we tackled the issue of node clustering in a graph under differential privacy constraints. This work is mostly related to differential privacy for structured datasets, especially in the specific framework of weighted graphs, in which sensitive informations are carried by the graph weights. 
We firstly introduced a new definition of differential privacy on graphs closely related to a "graph sketching framework" detailed in Section~\ref{section:NewPrivacy}.
Then, based on this new definition, we presented an algorithm for the release of an approximated minimum spanning tree topology of a weighted graph under differential privacy constraints. This method is, to our knowledge, the first In-processing method to produce an approximated spanning tree topology of a graph. Moreover this method outperforms theoretically and experimentally the state of the art of the Post-processing methods "Graph based Laplace mechanism" as soon as the graph is sufficiently dense. The theoretical analysis we produce is also scalable and thus adaptable to other related frameworks. 
Finally, we introduced a new method for nodes clustering under differential privacy constraints. This is, to our knowledge, the first time a differentially private almost minimum spanning tree is used to ensure privacy in a clustering algorithm.
This new algorithm for private MST-based-clustering could pave the way to applications in genomics, proteomics, web monitorings,  under weight differential privacy. This kind of analysis should keep good privacy while restoring the user faith in his/her security. This is why, as a future work, we aim at finding some theoretical guarantees for the accuracy and the robustness of our differentially clustering method. We will also continue to investigate such methods on simulated and on real life datasets. \\

\begin{normalsize}
\relscale{0.9}
\bibliographystyle{alpha}
\bibliography{sample-bibliography}

\openright
\appendix

\chapter{Some additional material}

\section{Graph Theory Notations}
\label{appendix:graphnotation}

In this report we consider any undirected-weighted graph $G$ as split in two parts denoting its topology $\mathcal{G}$ and its weight function $w$. The topology of a graph is composed by edge and node sets respectively $E$ and $V$, one can thus denote $\mathcal{G}=(V,E)$. The weight function is defined as $w:= E \rightarrow \mathbb{R}^{+}$ mapping any edge in $E$ to its weight in the graph. One should also denote $\mathcal{W}_{E}$ the set of all possible weight functions mapping $E$ to weights in $\mathbb{R}^{+}$.
\begin{remark}
Since any weighted graph is characterize by its topology and weight function $G=(\mathcal{G},w)=(V,E,w)$, a none weighted graph reduces to its topology $G=\mathcal{G}$
\end{remark}
Now that the weighted graph is defined, let us introduce some simple definitions.
\begin{deff}
Let $G=(V,E,w)$ be an undirected graph.
Let $u,v \in V$ such that $e=(u,v) \in E$, then $e$ is said to be an incident edge to $u$ and to $v$.
\end{deff}
\begin{deff}
Let $G=(V,E,w)$ be an undirected graph.
$G$ is said to be simple if it does not contain any graph loop (i.e. any edge having the same starting and ending node as illustrated by figure~\ref{figure:loopandmultiple}) or multiple edges (two different edges connecting the two same nodes).
\end{deff}

\begin{figure}[!ht]
\centering
\includegraphics[width=0.9\textwidth]{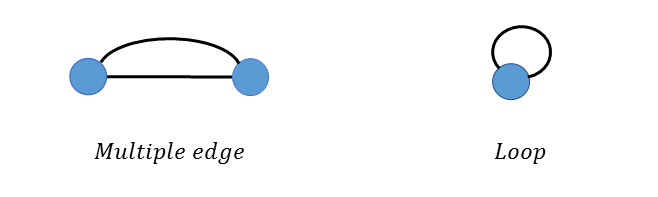}
\caption{Classical representation of a loop and a multiple edge}
\label{figure:loopandmultiple}
\end{figure}

For this work, every graph, even when not specified are supposed to be simple and undirected and connected.
Finally, since Chapter~\ref{section:PrivateMSTs} presents a Prim-like algorithm, it might be useful for any reader to look over this preliminary work in~\cite{6773228}, and the following definition closely linked to this algorithm.

\begin{deff}
Let $G=(V,E,w)$, and some set of vertices $S \subset V$, $e \in E$ is said to be xor-incident to $S$ if $e$ incident to one and only one vertex in $S$.
\end{deff}

\section{Proof of Prop 1.3.1}
\label{appendix:proofKulback}

\textbf{Recall of the proposition:}  A randomized algorithm $\mathcal{A}$ is $(\epsilon,\delta)$-differentially private if and only if for all $x,y \in \mathbb{N}^{|\mathcal{X}|}$ such that $x \sim y$,  $D^{\delta}_{\infty}\big(\mathcal{A}(x)||\mathcal{A}(y)\big) \leq \epsilon$ and $D^{\delta}_{\infty}\big(\mathcal{A}(y)||\mathcal{A}(x)\big) \leq \epsilon.$

\begin{preuve}
Let $\mathcal{A}$ be a randomized algorithm, and $x\sim y \in \mathbb{N}^{|\mathcal{X}|}$.\\
\fbox{$\implies$} If $\mathcal{A}$ is $(\epsilon,\delta)$ differential private. Then  $\forall  S \subset Range(\mathcal{A})$ one has 
\begin{align*}
&\mathbb{P}(\mathcal{A}(x) \in S) \leq e^{\epsilon}\mathbb{P}(\mathcal{A}(y) \in S) + \delta  \numberthis \label{equation:differentialprivacy}\\
\shortintertext{Since Eq.~(\ref{equation:differentialprivacy}) is true regardless of the value of $\mathbb{P}(\mathcal{A} \in S)$, it is also true when $S \subset Supp(\mathcal{A})$ and $\mathbb{P}(\mathcal{A}(x) \in S)>\delta$ then$ \forall S \subset Supp(\mathcal{A})$ such that $\mathbb{P}(\mathcal{A}(x) \in S)>\delta$ }
& \mathbb{P}(\mathcal{A}(x) \in S) \leq e^{\epsilon}\mathbb{P}(\mathcal{A}(y) \in S) + \delta \\
\implies & \mathbb{P}(\mathcal{A}(x) \in S) - \delta \leq e^{\epsilon}\mathbb{P}(\mathcal{A}(y) \in S) \\
\implies &  \ln\frac{\mathbb{P}(\mathcal{A}(x) \in S) - \delta}{\mathbb{P}(\mathcal{A}(y) \in S)}\leq \epsilon \\
\implies & D^{\delta}_{\infty}\big(\mathcal{A}(x)||\mathcal{A}(y)\big) \leq \epsilon\\
\shortintertext{Symmetrically one obtains $D^{\delta}_{\infty}\big(\mathcal{A}(y)||\mathcal{A}(x)\big) \leq \epsilon.$ } \notag
\end{align*}
\fbox{$\impliedby$} If $D^{\delta}_{\infty}\big(\mathcal{A}(x)||\mathcal{A}(y)\big) \leq \epsilon$ and  $D^{\delta}_{\infty}\big(\mathcal{A}(y)||\mathcal{A}(x)\big) \leq \epsilon$ then $\forall S \subset Range(\mathcal{A})$ 
\begin{itemize}
\item If $\mathbb{P}(\mathcal{A}(x) \in S) >\delta$ and $\mathbb{P}(\mathcal{A}(y) \in S) >0$ then \begin{align*}
\ln\left(\frac{\mathbb{P}(\mathcal{A}(x) \in S) - \delta}{\mathbb{P}(\mathcal{A}(y) \in S)}\right)\leq \epsilon \implies \mathbb{P}(\mathcal{A}(x) \in S) \leq e^{\epsilon}\mathbb{P}(\mathcal{A}(y) \in S) + \delta. 
\end{align*}

\item Else if $\mathbb{P}(\mathcal{A}(y) \in S) = 0$ or $\mathbb{P}(\mathcal{A}(x) \in S) < \delta$ the result is immediate.\hfill\ensuremath{\blacksquare}
\end{itemize}
\end{preuve}

\section{Complementary proof for theorem 4.2.1 }
\label{appendixcomplementaryprooflaplacecondition}

\textbf{Recall of the result:} Let $\epsilon >0$, $Y_{1},Y_{2} \overset{iid}\sim Lap(1/\epsilon)$, then $f_{Y_{1} + Y_{2}}(x) = \frac{\epsilon}{4}(1+ \epsilon|x|)e^{-\epsilon|x|}.$ 

\begin{preuve}

\begin{align*}
f_{Y_{1} + Y_{2}}(x)&=(f_{Y_{1}}*f_{Y_{2}})(x)=\int_{\mathbb{R}}f_{Y_{1}}(t)f_{Y_{2}}(x-t) dt =\int_{\mathbb{R}}\left(\frac{\epsilon}{2}\right)^{2}e^{-\epsilon(|t|+|x-t|)} dt\\
\shortintertext{If $x>0$ on has }
f_{Y_{1} + Y_{2}}(x)&=\left(\frac{\epsilon}{2}\right)^{2}\left(\int_{\mathbb{R}^{-}}e^{2\epsilon t - \epsilon x} dt + \int_0^x e^{-\epsilon x}dt \int_x^{+\infty} e^{-2\epsilon t + \epsilon x} dt\right)\\
&=\left(\frac{\epsilon}{2}\right)^{2}\left(\frac{e^{-\epsilon x}}{2 \epsilon} + x e^{-\epsilon x}+ \frac{e^{\epsilon x}}{2 \epsilon} e^{-2\epsilon x} \right)\\
&=\frac{\epsilon}{4}e^{-\epsilon x} + \frac{\epsilon^{2}}{4}x e^{-\epsilon x} \numberthis \label{equation:xsup}.
\shortintertext{If $x<0$ on has }
f_{Y_{1} + Y_{2}}(x)&=\left(\frac{\epsilon}{2}\right)^{2}\left(\int_{-\infty}^{x}e^{2\epsilon t - \epsilon x} dt + \int_x^0 e^{\epsilon x}dt \int_0^{+\infty} e^{-2\epsilon t + \epsilon x} dt\right)\\
&=\left(\frac{\epsilon}{2}\right)^{2}\left(\frac{e^{\epsilon x}}{ \epsilon} - x e^{\epsilon x}\right) \\
&= \frac{\epsilon}{4}e^{\epsilon x} - \frac{\epsilon^{2}}{4}x e^{\epsilon x} \numberthis\label{equation:xinf}.
\shortintertext{combining Eq.~(\ref{equation:xsup}) and Eq.~(\ref{equation:xinf}) one gets the result. \hfill\ensuremath{\blacksquare}}                    
\end{align*}
\end{preuve}
\textbf{Recall of the result:} Let $\epsilon >0$, $Y_{1},Y_{2} \overset{iid}\sim Lap(1/\epsilon)$, then $$\forall x \geq 0, F_{Y_{1} + Y_{2}}(x) = 1- e^{-\epsilon x}\left(\frac{1}{2}+\frac{\epsilon x}{4}\right).$$

\begin{preuve}
Let $x >0$, then 
\begin{align*}
\int_{-\infty}^{x} f_{Y_{1}+Y_{2}}(t) dt =& \int_{-\infty}^x \frac{\epsilon}{4}\left(1+\epsilon |t|) e^{-\epsilon|t|}\right) dt\\
=& \frac{\epsilon}{4}\left( \underset{A}{\underline{\int_{-\infty}^0 e^{\epsilon t}- \epsilon t e^{\epsilon t} dt}} + \underset{B}{\underline{\int_0^x e^{-\epsilon t} + \epsilon t e^{- \epsilon t} dt}}\right) \numberthis\label{equation:1}\\
\shortintertext{Let split $A$ and $B$}
A =& \frac{1}{\epsilon} \left[e^{\epsilon t} \right]_{-\infty}^0 - \int_{-\infty}^0 \epsilon e^{\epsilon t} dt \\
=&\frac{1}{\epsilon} -\left[t e^{\epsilon t} \right]_{-\infty}^0 + \int_{-\infty}^0 e^{\epsilon t} dt =\frac{2}{\epsilon}.\\
B =& -\frac{1}{\epsilon} \left[ e^{-\epsilon t}\right]_0^x + \int_0^x t \epsilon e^{-\epsilon t} dt \\
=& -\frac{1}{\epsilon}  e^{-\epsilon x} + \frac{1}{\epsilon}-\left[t e^{\epsilon t} \right]_{0}^x +\int_{0}^t e^{-\epsilon t} dt\\
=& \frac{2}{\epsilon} -\frac{2}{\epsilon}e^{-\epsilon x} -x e^{-\epsilon x}.
\shortintertext{It suffices to sum $A$ and $B$ in Eq.~(\ref{equation:1}) to find the result. \hfill\ensuremath{\blacksquare} }
\end{align*}
\end{preuve}

\section{JDSE submission}
\label{appendix:JDSE}
\begin{abstract}
We investigate the problem of nodes clustering in a graph representation of a dataset under privacy constraints. Our contribution is twofold. First we formally define the concept of differential privacy for graphs and give an  application setting where nodes clustering under privacy constraints allows for a secure analysis of the experiment. Then we propose a theoretically motivated method combining a sanitizing mechanism (such as Laplace or Gaussian mechanism) with a Minimum Spanning Tree (MST)-based clustering algorithm. It provides an accurate method for nodes clustering in a graph while keeping the sensitive information contained in the edges weights of the graph private. We provide some theoretical results on the robustness of the Kruskal minimum spanning tree construction for both sanitizing mechanisms. These results exhibit which conditions the graph's weights should respect in order to consider that the nodes form well separated clusters. The method has been experimentally evaluated on simulated data, and preliminary results show the good behavior of the algorithm while identifying well separated clusters. An extended experimental evaluation will be presented at the conference.
\end{abstract}

\newpage
\includepdf{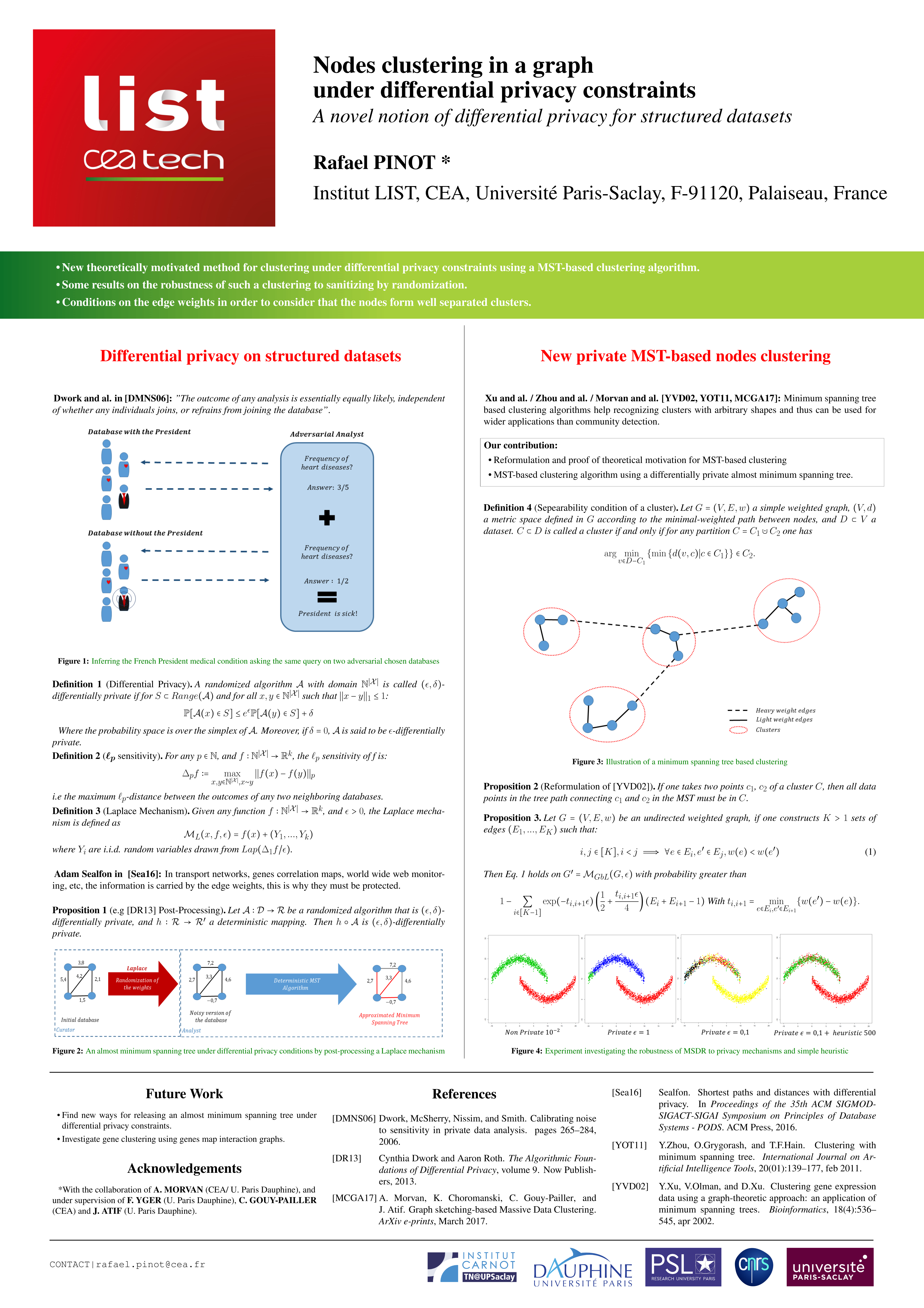}
\section{Additional Simulations}
\label{appendix:simu}
\begin{figure}[!ht]
\centering
\includegraphics[width=0.6\textwidth]{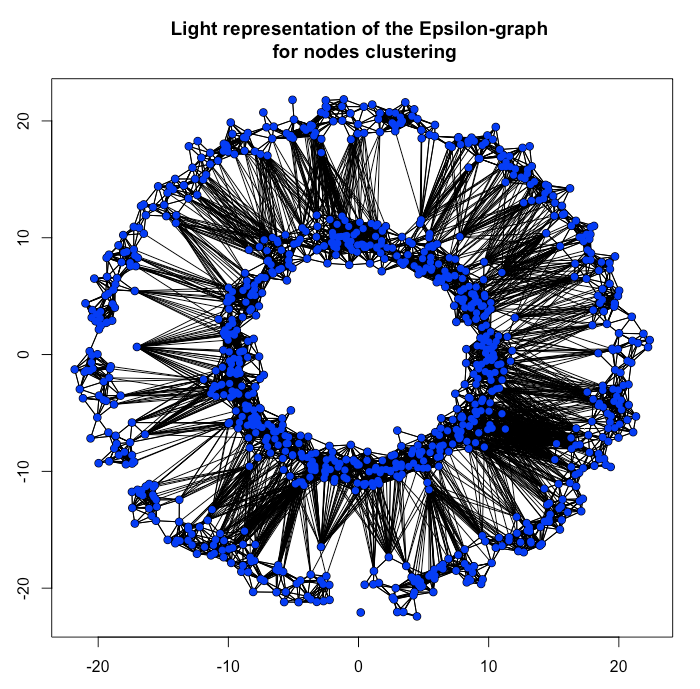}
\caption{Simple visualization of the $\epsilon$-graph based on the euclidean matrix distance of the simulated dataset 1}
\end{figure}

\begin{figure}[!ht]
\centering
\includegraphics[width=0.45\textwidth]{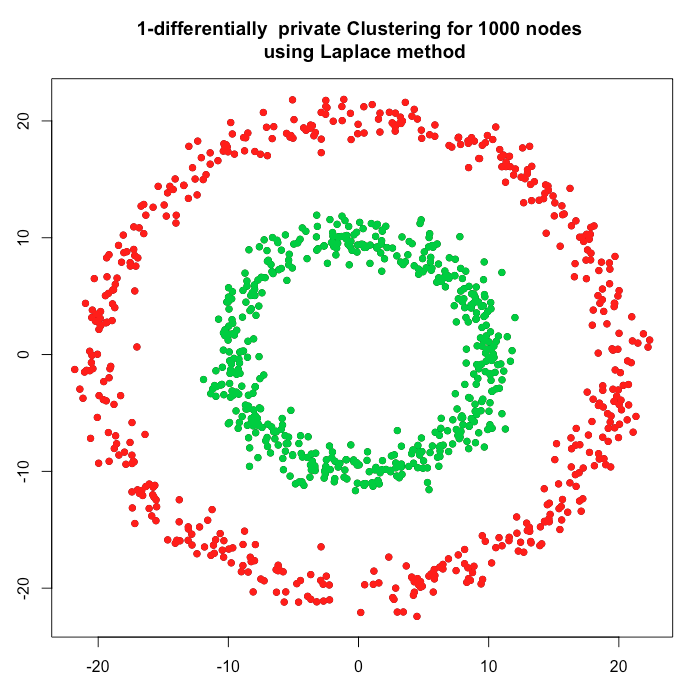}
\includegraphics[width=0.45\textwidth]{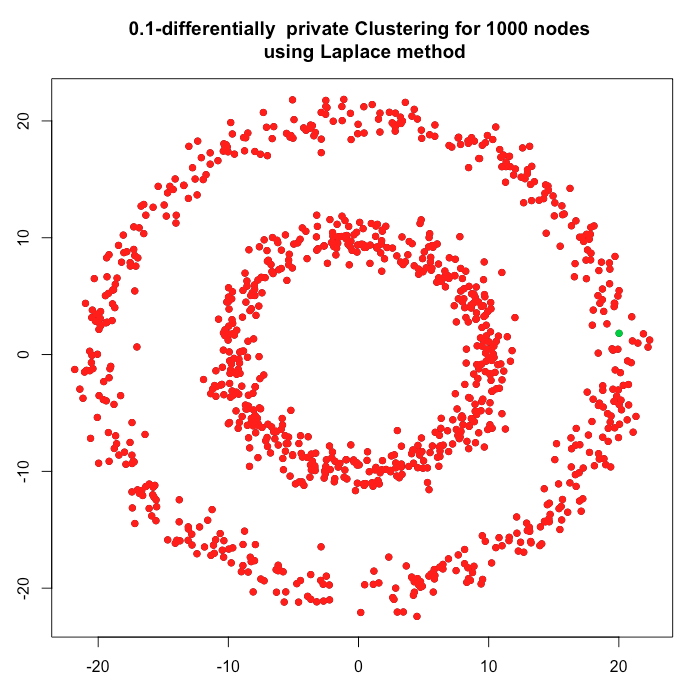}
\caption{Clustering under differential privacy on dataset 1 using Laplace sanitizing}
\end{figure}

\begin{figure}[t]
\centering
\includegraphics[width=0.45\textwidth]{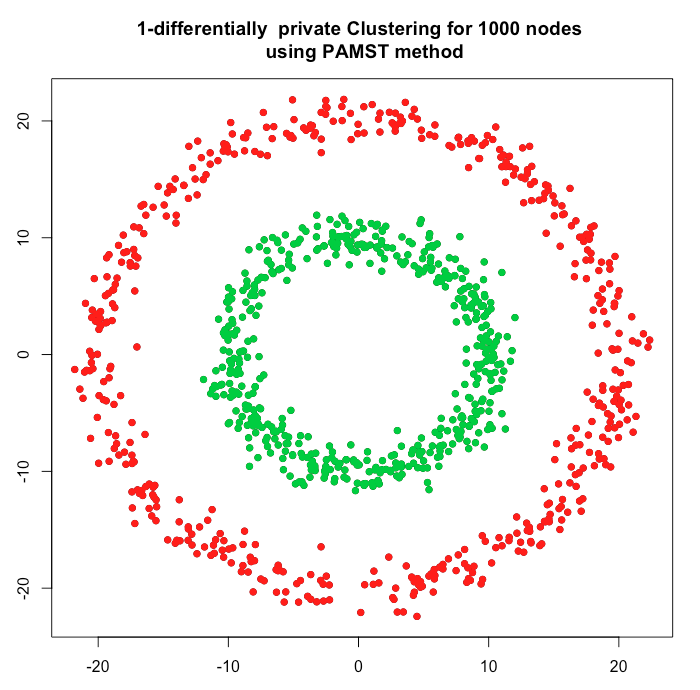}
\includegraphics[width=0.45\textwidth]{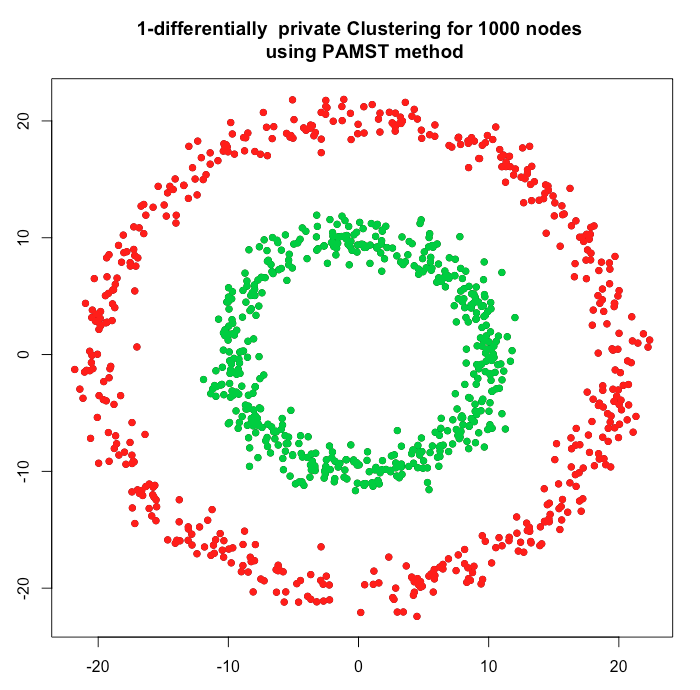}
\caption{Clustering under differential privacy on dataset 1 using \textsc{PAMST}}
\end{figure}

\begin{figure}[t]
\centering
\includegraphics[width=0.6\textwidth]{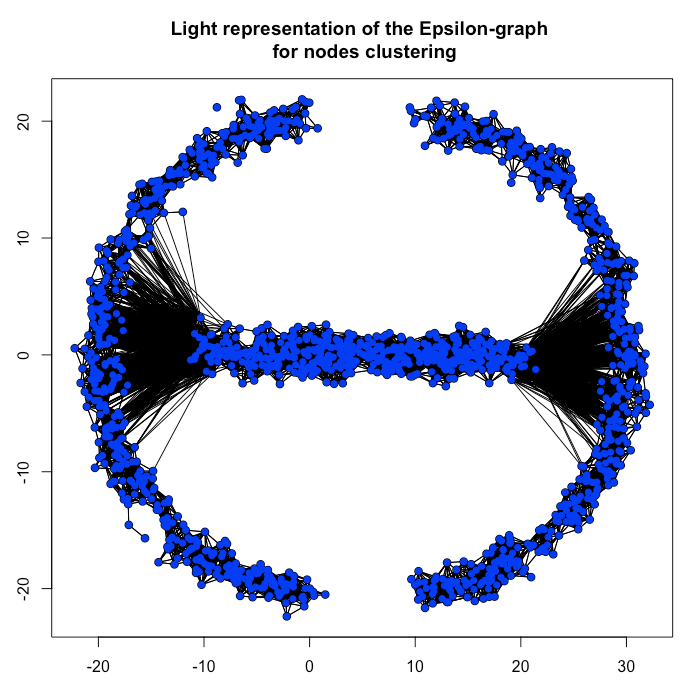}
\caption{Simple visualization of the $\epsilon$-graph based on the euclidean matrix distance of the dataset 2}
\end{figure}

\begin{figure}[t]
\centering
\includegraphics[width=0.45\textwidth]{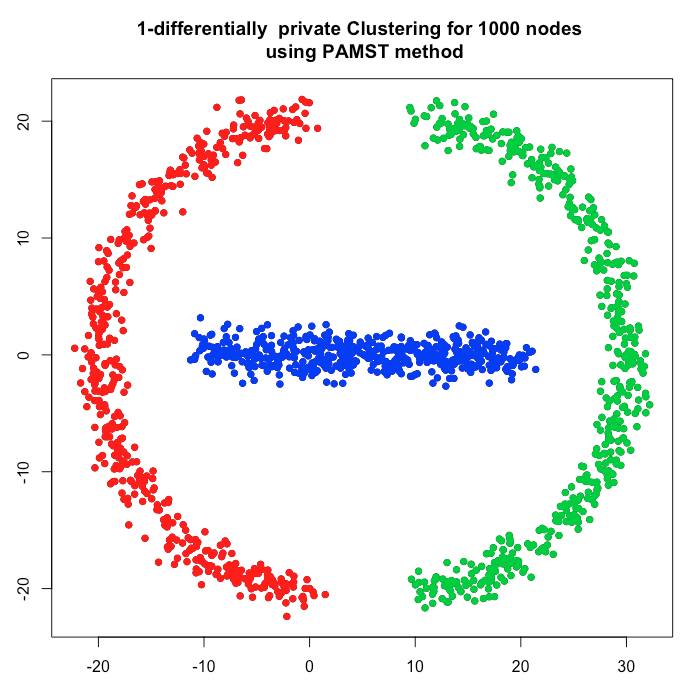}
\includegraphics[width=0.45\textwidth]{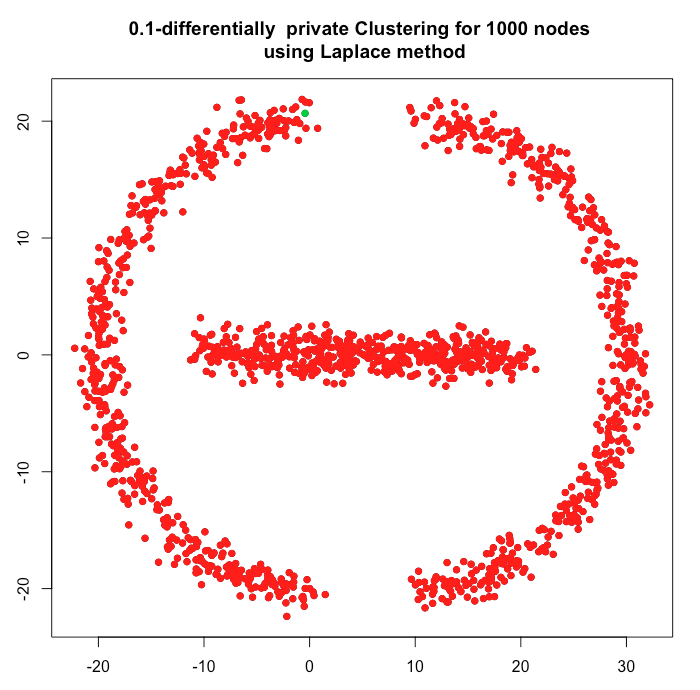}
\caption{Clustering under differential privacy on dataset 2 using Laplace sanitizing}
\end{figure}
        
\begin{figure}[t]
\centering
\includegraphics[width=0.45\textwidth]{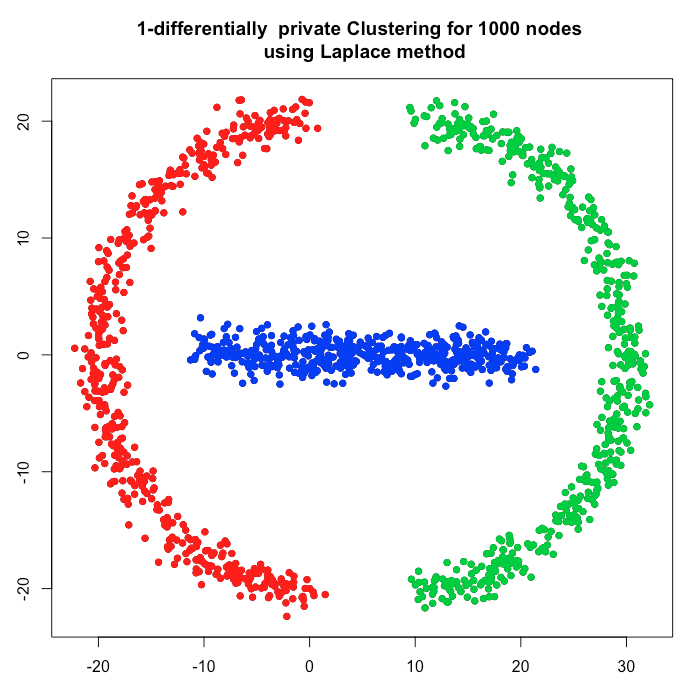}
\includegraphics[width=0.45\textwidth]{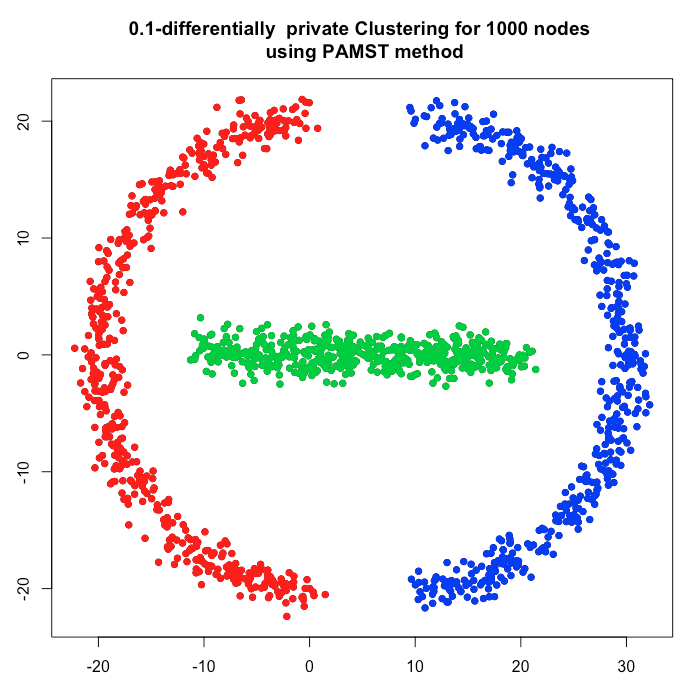}
\caption{Clustering under differential privacy on dataset 2 using \textsc{PAMST}}
\end{figure}
\end{normalsize}

\end{document}